%% file: IvNeuPsyDyn.tex
\documentclass[11pt]{article}
\usepackage{amssymb,amsfonts,amsmath,graphicx,hyperref}
\input{ivdiag}
\input{tcilatex}

\begin{document}

\title{Nonlinear Quantum Neuro--Psycho--Dynamics\\ with Topological Phase Transitions}
\author{Vladimir G. Ivancevic\thanks{%
Human Systems Integration, Land Operations Division, Defence
Science \& Technology Organisation, P.O. Box 1500, Edinburgh SA
5111, Australia
(Vladimir.Ivancevic@dsto.defence.gov.au)} \and Tijana T. Ivancevic\thanks{%
School of Electrical and Information Engineering, University of
South
Australia, Mawson Lakes, S.A. 5095, Australia (Tijana.Ivancevic@unisa.edu.au)%
}}
\date{}
\maketitle

\begin{abstract}
We have proposed a novel model of general quantum, stochastic and
chaotic psychodynamics. The model is based on the previously
developed Life--Space Foam (LSF) framework to motivational and
cognitive dynamics. The present model extends the LSF--approach by
incorporating chaotic and topological non-equilibrium phase
transitions. Such extended LSF--model is applied for rigorous
description of multi--agent joint action. The present model is
related to Haken--Kelso--Bunz model of self-organization in the
human motor system (including: multi-stability, phase transitions
and hysteresis effects, presenting a contrary view to the purely
feedback driven neural systems), as well as the entropy--approach
to adaptation in human goal--directed motor control.\\

\end{abstract}

\twocolumn

\section{Introduction}

Classical physics has provided a strong foundation for
understanding brain function through measuring brain activity,
modelling the functional connectivity of networks of neurons with
algebraic matrices, and modelling the dynamics of neurons and
neural populations with sets of coupled differential equations
(Freeman, 1975, 2000). Various tools from classical physics
enabled recognition and documentation of aspects of the physical
states of the brain; the structures and dynamics of neurons, the
operations of membranes and organelles that generate and channel
electric currents; and the molecular and ionic carriers that
implement the neural machineries of electrogenesis and learning.
They support description of brain functions at several levels of
complexity through measuring neural activity in the brains of
animal and human subjects engaged in behavioral exchanges with
their environments. One of the key properties of brain dynamics
are the coordinated oscillations of populations of neurons that
change rapidly in concert with changes in the environment (Freeman
and Vitiello, 2006; Ivancevic, 2006a, 2007b). Also, most
experimental neurobiologists and neural theorists have focused on
sensorimotor functions and their adaptations through various forms
of learning and memory. Reliance has been placed on measurements
of the rates and intervals of trains of action potentials of small
numbers of neurons that are tuned to perceptual invariances and
modelling neural interactions with discrete networks of simulated
neurons. These and related studies have given a vivid picture of
the cortex as a mosaic of modules, each of which performs a
sensory or motor function; they have not given a picture of
comparable clarity of the integration of modules.

According to Freeman and Vitiello (2006), \emph{many--body quantum
field theory} appears to be the only existing theoretical tool
capable to explain the dynamic origin of long--range correlations,
their rapid and efficient formation and dissolution, their interim
stability in ground states, the multiplicity of coexisting and
possibly noninterfering ground states, their degree of ordering,
and their rich textures relating to sensory and motor facets of
behaviors. It is historical fact that many--body quantum field
theory has been devised and constructed in past decades exactly to
understand features like ordered pattern formation and phase
transitions in condensed matter physics that could not be
understood in classical physics, similar to those in the brain.

The domain of validity of the `quantum' is not restricted to the
microscopic world (Umeza-va, 1993). There are macroscopic features
of classically behaving systems, which cannot be explained without
recourse to the quantum dynamics. This field theoretic model leads
to the view of the phase transition as a condensation that is
comparable to the formation of fog and rain drops from water
vapor, and that might serve to model both the gamma and beta phase
transitions. According to such a model, the production of activity
with long--range correlation in the brain takes place through the
mechanism of spontaneous breakdown of symmetry (SBS), which has
for decades been shown to describe long-range correlation in
condensed matter physics. The adoption of such a field theoretic
approach enables modelling of the whole cerebral hemisphere and
its hierarchy of components down to the atomic level as a fully
integrated macroscopic quantum system, namely as a macroscopic
system which is a quantum system not in the trivial sense that it
is made, like all existing matter, by quantum components such as
atoms and molecules, but in the sense that some of its macroscopic
properties can best be described with recourse to quantum dynamics
(see Freeman and Vitiello, 2006 and references therein).

It is well--known that non-equilibrium phase transitions (Haken,
1983, 1993, 1996) are phenomena which bring about {qualitative}
physical changes at the macroscopic level in presence of the same
microscopic forces acting among the constituents of a system.
Phase transitions can also be associated with autonomous robot
competence levels, as informal specifications of desired classes
of behaviors for robots over all environments they will encounter,
as described by Brooks' subsumption architecture approach (Brooks,
1986, 1989, 1990). The distributed network of augmented
finite--state machines can exist in different phases or modalities
of their state--space variables, which determine the systems
intrinsic behavior. The phase transition represented by this
approach is triggered by either internal (a set--point) or
external (a command) control stimuli, such as a command to
transition from a sleep mode to awake mode, or walking to running.

On the other hand, it is well--known that humans possess more
degrees of freedom than are needed to perform any defined motor
task, but are required to co-ordinate them in order to reliably
accomplish high-level goals, while faced with intense motor
variability. In an attempt to explain how this takes place,
Todorov and Jordan (2002) formulated an alternative theory of
human motor coordination based on the concept of stochastic
optimal feedback control. They were able to conciliate the
requirement of goal achievement (e.g., grasping an object) with
that of motor variability (biomechanical degrees of freedom).
Moreover, their theory accommodates the idea that the human motor
control mechanism uses internal `functional synergies' to regulate
task--irrelevant (redundant) movement.

Until recently, research concerning sensory processing and
research concerning motor control have followed parallel but
independent paths. The partitioning of the two lines of research
in practice partly derived from and partly fostered a bipartite
view of sensorimotor processing in the brain -- that a
sensory/perceptual system creates a general purpose representation
of the world which serves as the input to the motor systems (and
other cognitive systems) that generate action/behavior as an
output. Recent results from research on vision in natural tasks
have seriously challenged this view, suggesting that the visual
system does not generate a general--purpose representation of the
world, but rather extracts information relevant to the task at
hand (Droll et al, 2005; Land and Hayhoe, 2001). At the same time,
researchers in motor control have developed an increasing
understanding of how sensory limitations and sensory uncertainty
can shape the motor strategies that humans employ to perform
tasks. Moreover, many aspects of the problem of sensorimotor
control are specific to the mapping from sensory signals to motor
outputs and do not exist in either domain in isolation. Sensory
feedback control of hand movements, coordinate transformations of
spatial representations and the influence of processing speed and
attention on sensory contributions to motor control are just a few
of these. In short, to understand how human (and animal) actors
use sensory information to guide motor behavior, we must study
sensory and motor systems as an integrated whole rather than as
decomposable modules in a sequence of discrete processing steps
(Knill et al, 2007).

Cognitive neuroscience investigations, inclu-ding fMRI studies of
human co--action, suggest that cognitive and neural processes
supporting co--action include joint attention, action observation,
task sharing, and action coordination (Fogassi et al, 2005;
Knoblich and Jordan, 2003; Newman et al, 2007; Sebanz at al,
2006). For example, when two actors are given a joint control task
(e.g., tracking a moving target on screen) and potentially
conflicting controls (e.g., one person in charge of acceleration,
the other -- deceleration), their joint performance depends on how
well they can anticipate each other's actions. In particular,
better coordination is achieved when individuals receive
real--time feedback about the timing of each other's actions
(Sebanz at al, 2006).

A developing field in coordination dynamics involves the theory of
social coordination, which attempts to relate the DC to normal
human development of complex social cues following certain
patterns of interaction. This work is aimed at understanding how
human social interaction is mediated by meta-stability of neural
networks. fMRI and EEG are particularly useful in mapping
thalamocortical response to social cues in experimental studies.
In particular, a new theory called the \emph{Phi complex} has been
developed by S. Kelso and collaborators, to provide experimental
results for the theory of social coordination dynamics (see the
recent nonlinear dynamics paper discussing social coordination and
EEG dynamics of Tognoli et al, 2007). According to this theory, a
pair of phi rhythms, likely generated in the mirror neuron system,
is the hallmark of human social coordination. Using a dual-EEG
recording system, the authors monitored the interactions of eight
pairs of subjects as they moved their fingers with and without a
view of the other individual in the pair.

Recently developed Life Space Foam (LSF) model (Ivancevic and
Aidman, 2007) is an integration of two modern approaches to
cognition: (i) dynamical field theory (DFT, Amari, 1977;
Sch\"{o}ner, 2007) and (ii) quantum--probabilistic dynamics (QP,
Glimcher, 2005; Busemayer et al, 2006). In this paper we expand
the LSF--concept to model decision making process in human--robot
joint action and related LSF--phase transitions.

\section{Classical versus Quantum\\ Probability}

As \emph{quantum probability} in human cognition and decision
making has recently become popular, let us briefly describe this
fundamental concept (for more details, see Ivancevic, 2007a,
2007c, 2008b).

\subsection{Classical Probability and\\ Stochastic Dynamics}

Recall that a \textit{random variable} $X$ is defined by its
\textit{distribution function} $f(x)$. Its \emph{probabilistic
description} is based on the following rules: (i) $P(X=x_i)$ is
the probability that $X=x_i$; and (ii) $P(a\leq X\leq b)$ is the
probability that $X$ lies in a closed interval $[a,b]$. Its
statistical description is based on: (i) $\mu_X$ or $E(X)$ is the
mean or expectation of $X$; and (ii) $\sigma_X$ is the standard
deviation of $X$. There are two cases of random variables:
discrete and continuous, each having its own probability (and
statistics) theory.

A discrete random variable $X$ has only a countable number of values $%
\{x_{i}\}$. Its distribution function $f(x_{i})$ has the following
properties:
\begin{eqnarray*}
&&P(X=x_{i})=f(x_{i}),\qquad f(x_{i})\geq 0,\\
&&\sum_{i}f(x_{i})\,dx=1.
\end{eqnarray*}

Statistical description of $X$ is based on its discrete mean value
$\mu _{X}$ and standard deviation $\sigma _{X}$, given
respectively by
\begin{eqnarray*}
\mu _{X}&=&E(X)=\sum_{i}x_{i}f(x_{i}),\\ \sigma
_{X}&=&\sqrt{E(X^{2})-\mu _{X}^{2}}.
\end{eqnarray*}

Here $f(x)$ is a piecewise continuous function such that:
\begin{eqnarray*}
&&P(a\leq X\leq b)=\int_{a}^{b}f(x)\,dx,\qquad f(x)\geq 0,\\
&&\int_{-\infty }^{\infty }f(x)\,dx=\int_{\mathbb{R}}f(x)\,dx=1.
\end{eqnarray*}

Statistical description of $X$ is based on its continuous mean
$\mu _{X}$ and standard deviation $\sigma _{X}$, given
respectively by
\begin{eqnarray*}
&&\mu _{X}=E(X)=\int_{-\infty }^{\infty }xf(x)\,dx,\\ &&\sigma _{X}=\sqrt{%
 E(X^{2})-\mu _{X}^{2}}.
\end{eqnarray*}

Now, let us observe the similarity between the two descriptions.
The same kind of similarity between discrete and continuous
quantum spectrum stroke P. Dirac when he suggested the combined
integral approach, that he denoted by $\put(0,0){\LARGE
$\int$}\put(10,15){\small $\bf\Sigma$}\quad\,$ -- meaning `both
integral and sum at once': summing over a discrete spectrum and
integration over a continuous spectrum.

To emphasize this similarity even further, as well as to set--up
the stage for the path integral, recall the notion of a
\textit{cumulative distribution function} of a random variable
$X$, that is a function $F:{\mathbb R}\to{\mathbb R}$, defined by
$$F(a)=P(X)\leq a.$$ In particular, suppose that $f(x)$ is the
distribution function of $X$. Then
\begin{eqnarray*}
F(x)&=&\sum_{x_{i}\leq x}f(x_{i}),\qquad \text{or}\\
F(x)&=&\int_{-\infty }^{\infty }f(t)\,dt,
\end{eqnarray*}
according to as $x$ is a discrete or continuous random variable.
In either case, $F(a)\leq F(b)$  whenever $a\leq b$. Also, $$
\lim_{x\to -\infty}F(x)=0\qquad \text{and}\qquad \lim_{x\to
\infty}F(x)=1,
$$ that is, $F(x)$ is monotonic and its limit to the left is $0$
and the limit to the right is $1$. Furthermore, its cumulative
probability is given by
$$P(a\leq X\leq b)=F(b)-F(a),$$ and the Fundamental Theorem of
Calculus tells us that, in the continuum case,
$$f(x)=\partial_xF(x).$$

Now, recall that \textit{Markov stochastic process} is a random
process characterized by a \emph{lack of memory}, i.e., the
statistical properties of the immediate future are uniquely
determined by the present, regardless of the past (Gardiner, 1985;
Ivancevic, 2006b).

For example, a \textit{random walk} is an example of the
\textit{Markov chain}, i.e., a discrete--time Markov process, such
that the motion of the system in consideration is viewed as a
sequence of states, in which the transition from one state to
another depends only on the preceding one, or the probability of
the system being in state $k$ depends only on the previous state
$k-1$. The property of a Markov chain of prime importance in
biomechanics is the existence of an \emph{invariant distribution of states}%
: we start with an initial state $x_{0}$ whose absolute
probability is $1$. Ultimately the states should be distributed
according to a specified distribution.

Between the pure deterministic dynamics, in which all DOF of the
system in consideration are explicitly taken into account, leading
to classical dynamical equations, for example in Hamiltonian form
(using $\partial _{x}\equiv \partial /\partial x$),
\begin{equation}
\dot{q}^{i}=\partial _{p_{i}}H,\qquad \dot{p}_{i}=-\partial
_{q^{i}}H, \label{Ham1}
\end{equation}
(where $q^{i},p_{i}$ are coordinates and momenta, while $H=H(q,p)$
is the total system energy) -- and pure stochastic dynamics
(Markov process), there is so--called \emph{hybrid dynamics},
particularly \textit{Brownian dynamics}, in which some of DOF are
represented only through their \emph{stochastic influence} on
others. As an example, suppose a system of particles interacts
with a viscous medium. Instead of specifying a detailed
interaction of each particle with the particles of the viscous
medium, we represent the medium as a \emph{stochastic force}
acting on the particle. The stochastic force \emph{reduces the
dimensionally} of the dynamics.

Recall that the Brownian dynamics represents the phase--space
trajectories of a collection of particles that individually obey
\emph{Langevin rate equations} in the field of force (i.e., the
particles interact with each other via some deterministic force).
For a free particle, the Langevin equation reads (Gardiner, 1985):
\begin{eqnarray*}
m\dot{v}\,=\,R(t)\,-\,\beta v,
\end{eqnarray*}%
where $m$ denotes the mass of the particle and $v$ its velocity.
The right--hand side represent the coupling to a \emph{heat bath};
the effect of the random force $R(t)$ is to heat the particle. To
balance overheating (on the average), the particle is subjected to
\emph{friction} $\beta $. In humanoid dynamics this is performed
with the Rayleigh--Van der Pol's \emph{dissipation}. Formally, the
solution to the Langevin equation can be written as
\begin{eqnarray*}
&&v(t)\,=\,v(0)\,\exp\left( -\frac{\beta }{m}t\right)\\&&+\,\frac{1}{m}%
\int_{0}^{t}\exp[-(t-\tau )\beta /m]\,R(\tau )\,d\tau ,
\end{eqnarray*}%
where the integral on the right--hand side is a \textit{stochastic
integral} and the solution $v(t)$ is a random variable. The
stochastic properties of the solution depend significantly on the
stochastic properties of the random force $R(t)$. In the Brownian
dynamics the random force $R(t)$ is Gaussian distributed. Then the
problem boils down to finding the solution to the Langevin
stochastic differential equation with the supplementary condition
(zero and mean variance)
\begin{eqnarray*}
<R(t)>\,=\,0,\quad~ <R(t)\,R(0)>\,=\,2\beta k_{B}T\delta (t),
\end{eqnarray*}%
where $<.>$ denotes the mean value, $T$ is temperature,
$k_{B}-$\emph{equipartition} (i.e., uniform distribution of
energy) coefficient, Dirac $\delta (t)-$func-tion.

Algorithm for computer simulation of the Brownian dynamics (for a
single particle) can be written as (Heermann, 1990):
\begin{enumerate}
  \item Assign an initial position and velocity.
  \item Draw a random number from a Gaussian distribution with mean zero and
variance.
  \item Integrate the velocity to get $v^{n+1}$.
  \item Add the random component to the velocity.
\end{enumerate}

Another approach to taking account the coupling of the system to a
heat bath is to subject the particles to collisions with
\emph{virtual particles} (Heermann, 1990). Such collisions are
imagined to affect only momenta of the particles, hence they
affect the kinetic energy and introduce fluctuations in the total
energy. Each stochastic collision is assumed to be an
instantaneous event affecting only one particle.

The collision--coupling idea is incorporated into the Hamiltonian
model of dynamics (\ref{Ham1}) by adding a stochastic force
$R_{i}=R_{i}(t)$ to the $\dot{p}$ equation
\begin{eqnarray*}
\dot{q}^{i}=\partial _{p_{i}}H,\qquad \dot{p}_{i}=-\partial
_{q^{i}}H+R_{i}(t).
\end{eqnarray*}

On the other hand, the so--called \textit{Ito stochastic integral}
represents a kind of classical Riemann--Stieltjes integral from
linear functional analysis, which is (in $1$D case) for an
arbitrary time--function $G(t)$ defined as the \emph{mean square
limit}
\begin{eqnarray*}
&&\int_{t_{0}}^{t}G(t)dW(t)=\\ &&ms\lim_{n\rightarrow \infty
}\{\sum_{i=1}^{n}G(t_{i-1}[W(t_{i})-W(t_{i-1}]\}.
\end{eqnarray*}

Now, the general $N$D Markov process can be defined by \emph{Ito}
stochastic differential equation (SDE),
\begin{eqnarray*}
dx_{i}(t) &=&A_{i}[x^{i}(t),t]dt+B_{ij}[x^{i}(t),t]\,dW^{j}(t), \\
x^{i}(0) &=&x_{i0},\qquad (i,j=1,\dots ,N)
\end{eqnarray*}%
or corresponding \emph{Ito stochastic integral equation}
\begin{eqnarray*}
&&x^{i}(t)=x^{i}(0)+\int_{0}^{t}ds\,A_{i}[x^{i}(s),s]\\&&+\int_{0}^{t}dW^{j}(s)%
\,B_{ij}[x^{i}(s),s],
\end{eqnarray*}%
in which $x^{i}(t)$ is the variable of interest, the vector
$A_{i}[x(t),t]$ denotes deterministic \emph{drift}, the matrix
$B_{ij}[x(t),t]$ represents
continuous stochastic \emph{diffusion fluctuations}, and $W^{j}(t)$ is an $%
N-$ variable \textit{Wiener process} (i.e., generalized Brownian
motion, see Wiener, 1961) and $$dW^{j}(t)=W^{j}(t+dt)-W^{j}(t).$$

Now, there are three well--known special cases of the
\textit{Chapman--Kolmogorov equation} (see Gardiner, 1985):
\begin{enumerate}
    \item When both $B_{ij}[x(t),t]$ and $W(t)$ are zero, i.e., in the case of
pure deterministic motion, it reduces to the \textit{Liouville
equation}
\begin{eqnarray*}
&&\partial _{t}P(x',t'|x'',t'')=\\&&-\sum_{i}\frac{\partial }
{\partial x^{i}}\left\{ A_{i}[x(t),t]\,P(x'%
,t'|x'',t'')\right\} .
\end{eqnarray*}
    \item When only $W(t)$ is zero, it reduces to the
    \textit{Fokker--Planck equation}
\begin{eqnarray*}
\partial _{t}P(x',t'|x'',t'')= \hspace{3cm}\\
-\sum_{i}\frac{\partial }{\partial x^{i}}\left\{ A_{i}[x(t),t]\,P(x'%
,t'|x'',t'')\right\} +\\
\frac{1}{2}\sum_{ij}\frac{\partial ^{2}}{\partial x^{i}\partial
x^{j}}\left\{ B_{ij}[x(t),t]\,P(x',t'|x'',t'')\right\} .
\end{eqnarray*}
    \item When both $A_{i}[x(t),t]$ and $B_{ij}[x(t),t]$ are zero,
i.e., the state--space consists of integers only, it reduces to
the \textit{Master equation} of discontinuous jumps
\begin{eqnarray*}
&&\partial _{t}P(x',t'|x'',t'') =\\
&&\int dx\,W(x'|x'',t)\,P(x',t'|x'',t'')\\&&-\int dx\,W(x''|x',t)\,P(%
x',t'|x'',t'').
\end{eqnarray*}
\end{enumerate}

The \textit{Markov assumption} can now be formulated in terms of
the conditional probabilities $P(x^{i},t_{i})$: if the times
$t_{i}$ increase from right to left, the conditional probability
is determined entirely by the knowledge of the most recent
condition. Markov process is generated by a set of conditional
probabilities whose probability--density $P=P(x',t'|x'',t'')$
evolution obeys the general \textit{Chapman--Kolmogorov
integro--differen-tial equation}
\begin{eqnarray*}
&&\partial _{t}P =-\sum_{i}\frac{\partial }{\partial x^{i}}\left\{
A_{i}[x(t),t]\,P\right\} \qquad  \\
&&+\frac{1}{2}\sum_{ij}\frac{\partial ^{2}}{\partial x^{i}\partial x^{j}}%
\left\{ B_{ij}[x(t),t]\,P\right\}\\ &&+\int dx\left\{ W(x^{\prime
}|x^{\prime \prime },t)\,P-W(x^{\prime \prime }|x^{\prime
},t)\,P\right\}
\end{eqnarray*}
including \emph{deterministic drift}, \emph{diffusion
fluctuations} and \emph{discontinuous jumps} (given respectively
in the first, second and third terms on the r.h.s.).

It is this general Chapman--Kolmogorov inte-gro--differential
equation, with its conditional probability density evolution,
$P=P(x',t'|x'',t'')$, that we are going to model by the \emph{Feynman path integral} $\put(0,0){\LARGE $\int$%
}\put(10,15){\small $\bf\Sigma$}\quad\,$, providing us with the
physical insight behind the abstract (conditional) probability
densities.

\subsection{Quantum Probability Concept}

An alternative concept of probability, the so--called
\emph{quantum probability}, is based on the following physical
facts (elaborated in detail in this section):
\begin{enumerate}
    \item \emph{The time--dependent Schr\"{o}dinger equation} represents a
\emph{complex--valued generalization} of the real--valued
\textit{Fokker--Planck equation} for describing the
spatio--temporal \emph{probability density function} for the
system exhibiting \emph{continuous--time Markov stochastic
process}.
    \item The \emph{Feynman path integral} $\put(0,0){\LARGE $\int$%
}\put(10,15){\small $\bf\Sigma$}\quad\,$ is a generalization of
the time--dependent Schr\"{o}-dinger equation, including both
continuous--time and discrete--time Markov stochastic processes.
    \item Both Schr\"{o}dinger equation and path integral give `physical description' of
    any system they are modelling in terms of its physical energy, instead of an abstract probabilistic
description of the Fokker--Planck equation.
\end{enumerate}

Therefore, the \textit{Feynman path integral} $\put(0,0){\LARGE $\int$%
}\put(15,15){\small $\bf\Sigma$}\quad\,$, as a generalization of
the time--dependent Schr\"{o}-dinger equation, gives a unique
physical description for the general Markov stochastic process, in
terms of the physically based generalized probability density
functions, valid both for continuous--time and discrete--time
Markov systems.

{\it Basic consequence: a different way for calculating
probabilities.} The difference is rooted in the fact that
\textsl{sum of squares is different from the square of sums}, as
is explained in the following text.

\begin{figure}[h]
\centerline{\includegraphics[height=5cm]{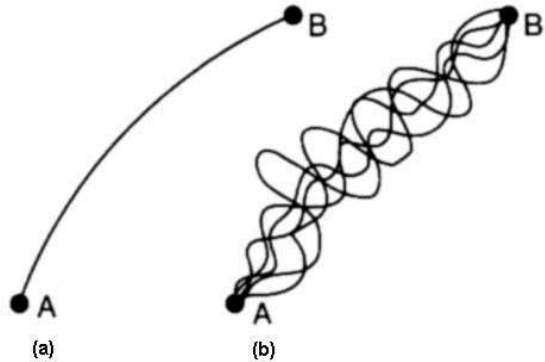}}
\caption{{\protect\small {Two ways of physical \emph{transition}
from an \emph{initial state} $A$ to the corresponding \emph{final
state} $B$. $\quad$ (a) Classical physics proposes \emph{a single
deterministic trajectory}, minimizing the total system's energy.
(b) Quantum physics proposes \emph{a family of Markov stochastic
histories}, namely \emph{all possible routes} from $A$ to $B$,
both continuous--time and discrete--time Markov chains, each
giving an equal contribution to the total \emph{transition
probability}.}}} \label{Paths}
\end{figure}

In Dirac--Feynman quantum formalism, each possible route from the
initial system state $A$ to the final system state $B$ is called a
\emph{history}. This history comprises any kind of a route (see
Figure \ref{Paths}), ranging from continuous and smooth
deterministic (mechanical--like) paths to completely discontinues
and random Markov chains (see e.g., Gardiner, 1985).
Each history (labelled by index $k$) is quantitatively described by a \emph{%
complex number,}$\ z_{k}=r_{k}\mathrm{e}^{\mathrm{i}\theta _{k}}$
(with $\mathrm{i}=\sqrt{-1}$), called the `individual transition
amplitude'. Its absolute square, $|z_{k}|^{2}$, is called the
\emph{individual transition probability}. Now, the \emph{total
transition amplitude} is the sum of all individual transition amplitudes, $%
\sum_{k}z_{k}$, called the \emph{sum--over--histo-ries}. The
absolute square of this sum--over--histories,
$|\sum_{k}z_{k}|^{2}$, is the \emph{total transition probability}.

In this way, the overall probability of the system's transition
from some initial state $A$ to some final state $B$ is given
\emph{not} by adding up the probabilities for each history--route,
but by `head--to--tail' adding up the sequence of amplitudes
making--up each route first (i.e., performing the
sum--over--histories) -- to get the total amplitude as a
`resultant vector', and then squaring the total amplitude to get
the overall transition probability.

\section{The Life Space Foam}

General nonlinear attractor dynamics, both deterministic and
stochastic, as well as possibly chaotic, developed in the
framework of Feynman path integrals, have recently been applied by
Ivancevic and Aidman (2007) to formalize classical Lewinian
field--theoretic psychodynamics (Lewin, 1951, 1997; Gold, 1999),
resulting in the development of a new concept of {life--space
foam} (LSF) as a natural medium for motivational and cognitive
psychodynamics. According to the LSF--formalism, the classic
Lewinian life space can be macroscopically represented as a smooth
manifold with steady force--fields and behavioral paths, while at
the microscopic level it is more realistically represented as a
collection of wildly fluctuating force--fields, (loco)mo-tion
paths and local geometries (and topologies with holes).

\begin{figure}[tbh]
\centerline{\includegraphics[width=13cm]{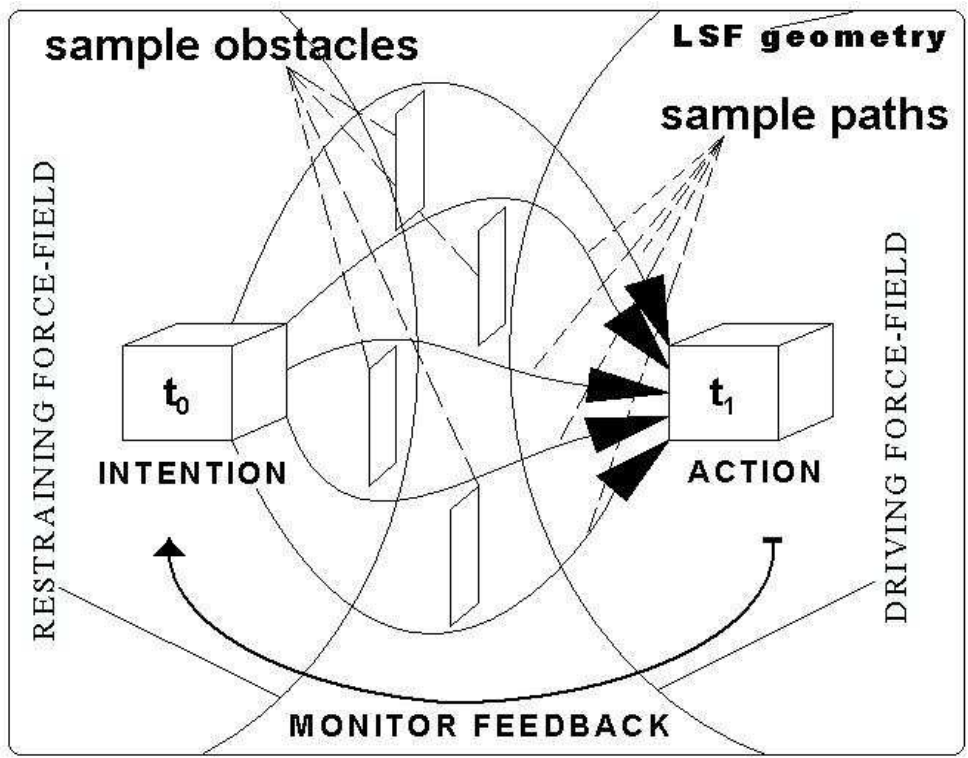}}
\caption{\textit{Transition--propagator} corresponding to each of
the motivational phases
$\{\mathcal{F},\mathcal{I},\mathcal{M},\mathcal{T}\}$, consisting
of an ensemble of feedforward paths propagating through the `wood
of obstacles'. The paths affected by driving and restraining
force--fields, as well as by the local LSF--geometry. Transition
goes from $Intention$, occurring at a sample time instant $t_{0}$,
to $Action$, occurring at some later time $t_{1}$. Each propagator
is controlled by its own $Monitor$ feedback.} \label{IvPaths}
\end{figure}

We have used the new LSF concept to develop modelling framework
for motivational dynamics (MD) and induced cognitive dynamics
(CD). Motivation processes both precede and coincide with every
goal--directed action. Usually these motivation processes include
the sequence of the following four feedforward \textit{phases}
(Ivancevic and Aidman, 2007): (*)
\begin{enumerate}
    \item \textit{Intention Formation} $\mathcal{F}$, including: decision making, commitment building, etc.
    \item \textit{Action Initiation} $\mathcal{I}$, including: handling conflict of motives, resistance to
    alternatives, etc.
    \item \textit{Maintaining the Action} $\mathcal{M}$, including: resistance to fatigue, distractions, etc.
    \item \textit{Termination} $\mathcal{T}$, including parking and\\ avoiding addiction, i.e., staying in
    control.
\end{enumerate}
With each of the phases
$\{\mathcal{F},\mathcal{I},\mathcal{M},\mathcal{T}\}$ in (*), we
can associate a \textit{transition propagator} -- an ensemble of
(possibly crossing) feedforward paths propagating through the
`wood of obstacles' (including topological holes in the LSF, see
Figure \ref{IvPaths}), so that the complete transition is a
product of propagators (as well as sum over paths). All the
phases--propagators are controlled by a unique $Monitor$ feedback
process.

A set of least--action principles is used to model the smoothness
of global, macro--level LSF paths, fields and geometry, according
to the following prescription. The action $S[\Phi]$,
psycho--physical dimensions of $$Energy\times Time=\text{\it
Effort}$$ ~~and depending on macroscopic paths, fields and
geometries (commonly denoted by an abstract field symbol $\Phi^i$)
is defined as a temporal integral from the {initial} time instant
$t_{ini}$ to the {final} time instant $t_{fin}$,
\begin{equation}
S[\Phi]=\int_{t_{ini}}^{t_{fin}}\mathfrak{L}[\Phi]\,dt,
\label{act}
\end{equation}%
with {Lagrangian density} given by
\begin{eqnarray*}
\mathfrak{L}[\Phi]=\int
d^{n}x\,\mathcal{L}(\Phi_i,\partial_{x^j}\Phi^i),
\end{eqnarray*}%
where the integral is taken over all $n$ coordinates $x^j=x^j(t)$
of the LSF, and $\partial_{x^j}\Phi^i$ are time and space partial
derivatives of the $\Phi^i-$variables over coordinates. The
standard {least action principle}
\begin{equation}
\delta S[\Phi]=0, \label{actPr}
\end{equation}
gives, in the form of the so--called Euler--Lagran-gian equations,
a shortest (loco)motion path, an extreme force--field, and a
life--space geometry of minimal curvature (and without holes). In
this way, we effectively derive a {unique globally smooth
transition map}
\begin{equation}
F\;:\;INTENTION_{t_{ini}}\cone{} ACTION_{t_{fin}}, \label{unique}
\end{equation}
performed at a macroscopic (global) time--level from some initial
time $t_{ini}$ to the final time $t_{fin}$. In this way, we have
obtained macro--objects in the global LSF: a single path described
by Newtonian--like equation of motion, a single force--field
described by Maxwellian--like field equations, and a single
obstacle--free Riemannian geometry (with global topology\\ without
holes).

To model the corresponding local, micro--level LSF structures of
rapidly fluctuating cognitive dynamics, an adaptive path integral
is formulated, defining a multi--phase and multi--path
(multi--field and multi--geometry) {transition amplitude} from the
state of $Intention$ to the state of $Action$,
\begin{equation}
\langle Action|Intention\rangle_{total} :=\put(0,0){\LARGE
$\int$}\put(15,15){\small $\bf\Sigma$}\quad\,\mathcal{D}[w\Phi]\,
{\mathrm e}^{\mathrm i S[\Phi]}, \label{pathInt1}
\end{equation}
where the Lebesgue integration is performed over all continuous
$\Phi^i_{con}=paths+fields+geometries$, while summation is
performed over all discrete processes and regional topologies
$\Phi^j_{dis}$. The symbolic differential $\mathcal{D}[w\Phi]$ in
the general path integral (\ref{pathInt1}), represents an
{adaptive path measure}, defined as a weighted product (with
${i=1,...,n=con+dis}$)
\begin{equation}
\mathcal{D}[w\Phi]=\lim_{N\to\infty}\prod_{s=1}^{N}w_sd\Phi
_{s}^i. \label{prod1}
\end{equation}
The adaptive path integral (\ref{pathInt1})--(\ref{prod1})
represents an $\infty-$dimensional neural network, with weights
$w$ updating by the general rule (Ivancevic and Aidman, 2007):
$$new\;value(t+1) = old\;value(t)\,+\, innovation(t).$$

The adaptive path integral (\ref{pathInt1}) incorporates the local
Bernstein adaptation process (Bernstein, 1967, 1982):
\begin{eqnarray*}
desired\;state~SW(t+1)\;=
\hspace{3cm}\\current\;state~IW(t)+adjustment~step~\Delta W(t)
\end{eqnarray*}
as well as the {augmented finite state machine} of Brooks'
subsumption architecture (Brooks, 1986, 1989, 1990), with a
networked behavior function:
\begin{eqnarray*}
&&final\;state~w(t+1)\; =\\ &&current\;state~w(t)\,+\\
&&adjustment~behavior ~f(\Delta w(t)).
\end{eqnarray*}

We remark here that the traditional neural networks approaches are
known for their classes of functions they can represent. This
limitation has been attributed to their low-dimensionality (the
largest neural networks are limited to the order of $10^5$
dimensions, see Izhikevich and Edelman, 2008). The proposed path
integral approach represents a new family of
function-re-presentation methods, which potentially offers a basis
for a fundamentally more expansive solution.

On the macro--level in $LSF$ we have the (loco)\textit{motion
action principle}
$$
\delta S[x]=0,
$$
with the \textit{Newtonian--like action} $S[x]$ given by
\begin{equation}
S[x]=\int_{t_{ini}}^{t_{fin}}dt\,[{1\over2}g_{ij}\,\dot{x}^{i}\dot{x}^{j}+\varphi^i(x^i)],
\label{actmot}
\end{equation}
where $\dot{x}^i$ represents motivational (loco)motion velocity
vector with cognitive \textit{processing speed}. The first bracket
term in (\ref{actmot}) represents the kinetic energy $T$,
$$T={1\over2}g_{ij}\,\dot{x}^{i}\dot{x}^{j},$$ generated by the \textit{Riemannian
metric tensor} $g_{ij}$, while the second bracket term,
$\varphi^i(x^i)$, denotes the family of potential force--fields,
driving the (loco)motions $x^i=x^i(t)$ (the \textit{strengths} of
the fields $\varphi^i(x^i)$ depend on their positions $x^i$ in
LSF. The corresponding Euler--Lagrangian equation gives the
Newtonian--like equation of motion
\begin{equation}
\frac{d}{dt}T_{\dot{x}^{i}}-T_{x^{i}}=-\varphi^i_{x^i},
\label{Newton}
\end{equation}
(subscripts denote the partial derivatives), which can be put into
the standard Lagrangian form
$$\frac{d}{dt}L_{\dot{x}^{i}}=L_{x^{i}},\qquad\text{with}\qquad
L=T-\varphi^i(x^i).
$$

Now, according to Lewin, the life space also has a sophisticated
topological structure. As a Riemannian smooth $n-$manifold, the
LSF--manifold $\bf\Sigma$ gives rise to its fundamental $n-$
\emph{groupoid}, or $n-$category $\Pi _{n}(\Sigma)$ (see
Ivancevic, 2006b, 2007a). In $\Pi _{n}(\Sigma)$, 0--cells are
\emph{points} in $\bf\Sigma$; 1--cells are \emph{paths} in
$\bf\Sigma$ (i.e.,
parameterized smooth maps $f:[0,1]\rightarrow \Sigma$); 2--cells are \emph{%
smooth homotopies} (denoted by $\simeq $) \emph{of paths} relative
to endpoints (i.e., parameterized smooth maps $h:[0,1]\times
\lbrack 0,1]\rightarrow \Sigma$); 3--cells are \emph{smooth
homotopies of homotopies} of paths in $\bf\Sigma$ (i.e.,
parameterized smooth maps $j:[0,1]\times \lbrack 0,1]\times
\lbrack 0,1]\rightarrow \Sigma$). Categorical \emph{composition}
is defined by \emph{pasting} paths and homotopies. In this way,
the following \textit{recursive homotopy dynamics} emerges on the
LSF--manifold $\bf\Sigma$ (**):

\onecolumn

\label{ivncat}
\begin{eqnarray*}
&&\mathtt{0-cell:}\,\,x_{0}\,\node\,\,\,\qquad x_{0}\in M; \qquad
\text{in
the higher cells below: }t,s\in[0,1]; \\
&&\mathtt{1-cell:}\,\,x_{0}\,\node\cone{f}\node\,x_{1}\qquad
f:x_{0}\simeq
x_{1}\in M, \\
&&f:[0,1]\rightarrow M,\,f:x_{0}\mapsto
x_{1},\,x_{1}=f(x_{0}),\,f(0)=x_{0},\,f(1)=x_{1}; \\
&&\text{e.g., linear path: }f(t)=(1-t)\,x_{0}+t\,x_{1};\qquad \text{or} \\
&&\text{Euler--Lagrangian }f-\text{dynamics with endpoint conditions }%
(x_0,x_1): \\
&&\frac{d}{dt}f_{\dot{x}^{i}}=f_{x^{i}},\quad \text{with}\quad
x(0)=x_{0},\quad x(1)=x_{1},\quad (i=1,...,n); \\
&&\mathtt{2-cell:}\,\,x_{0}\,\node\ctwodbl{f}{g}{h}\node\,x_{1}\qquad
h:f\simeq g\in M, \\
&&h:[0,1]\times \lbrack 0,1]\rightarrow M,\,h:f\mapsto g,\,g=h(f(x_{0})), \\
&&h(x_{0},0)=f(x_{0}),\,h(x_{0},1)=g(x_{0}),\,h(0,t)=x_{0},\,h(1,t)=x_{1} \\
&&\text{e.g., linear homotopy: }h(x_{0},t)=(1-t)\,f(x_{0})+t\,g(x_{0});\qquad%
\text{or} \\
&&\text{homotopy between two Euler--Lagrangian
}(f,g)-\text{dynamics}
\\
&&\text{with the same endpoint conditions }(x_0,x_1): \\
&&\frac{d}{dt}f_{\dot{x}^{i}}=f_{x^{i}},\quad \text{and} \quad \frac{d}{dt}%
g_{\dot{x}^{i}}=g_{x^{i}}\quad\text{with}\quad x(0)=x_{0},\quad
x(1)=x_{1};
\\
&&\mathtt{3-cell:}\,\,x_{0}\,\node\cthreecelltrp{f}{g}{h}{i}{j}\node%
\,x_{1}\qquad j:h\simeq i\in M, \\
&&j:[0,1]\times \lbrack 0,1]\times \lbrack 0,1]\rightarrow
M,\,j:h\mapsto
i,\,i=j(h(f(x_{0}))) \\
&&j(x_{0},t,0)=h(f(x_{0})),\,j(x_{0},t,1)=i(f(x_{0})), \\
&&j(x_{0},0,s)=f(x_{0}),\,j(x_{0},1,s)=g(x_{0}), \\
&&j(0,t,s)=x_{0},\,j(1,t,s)=x_{1} \\
&&\text{e.g., linear composite homotopy: }j(x_{0},t,s)=(1-t)\,h(f(x_{0}))+t%
\,i(f(x_{0})); \\
&&\text{or, homotopy between two homotopies between above two Euler-} \\
&&\text{Lagrangian }(f,g)-\text{dynamics with the same endpoint conditions }%
(x_0,x_1).
\end{eqnarray*}

\twocolumn

On the micro--LSF level, instead of a single path defined by the
Newtonian--like equation of motion (\ref{Newton}), we have an
ensemble of fluctuating and crossing paths with weighted
probabilities (of the unit total sum). This ensemble of
micro--paths is defined by the simplest instance of our adaptive
path integral (\ref{pathInt1}), similar to the Feynman's original
\textit{sum over histories},
\begin{equation}
\langle Action|Intention\rangle_{paths}=\put(0,0){\LARGE
$\int$}\put(15,15){\small $\bf\Sigma$}\quad\, \mathcal{D}[wx]\,
\mathrm{e}^{{\mathrm i} S[x]},  \label{Feynman}
\end{equation}
where $\mathcal{D}[wx]$ is a functional measure on the
\textit{space of all weighted paths}, and the exponential depends
on the action $S[x]$ given by (\ref{actmot}). This procedure can
be redefined in a mathematically cleaner way if we Wick--rotate
the time variable $t$ to imaginary values, $t\mapsto \tau={\mathrm
i} t$, thereby making all integrals real:
\begin{equation}
\put(0,0){\LARGE $\int$}\put(15,15){\small $\bf\Sigma$}\quad\,
\mathcal{D}[wx]\, \mathrm{e}^{{\mathrm i} S[x]}~\cone{Wick}\quad
\put(0,0){\LARGE $\int$}\put(15,15){\small $\bf\Sigma$}\quad\,
\mathcal{D}[wx]\, \mathrm{e}^{-S[x]}. \label{Wick}
\end{equation}
Discretization of (\ref{Wick}) gives the standard
\textit{thermo-dynamic--like partition function}
\begin{equation}
Z=\sum_j{\mathrm e}^{-w_jE^j/T}, \label{partition}
\end{equation}
where $E^j$ is the motion energy eigenvalue (reflecting each
possible motivational energetic state), $T$ is the
temperature--like environmental control parameter, and the sum
runs over all motion energy eigenstates (labelled by the index
$j$). From (\ref{partition}), we can further calculate all
thermodynamic--like and statistical properties of MD and CD, as
for example, \textit{transition entropy}, $S = k_B\ln Z$, etc.

\subsection{Noisy Decision Making in the LSF}

From CD--perspective, our adaptive path integral (\ref{Feynman})
calculates all (alternative) pathways of information flow during
the transition $Intention\to Action$. In the connectionist
language, (\ref{Feynman}) represents \textit{activation dynamics},
to which our $Monitor$ process gives a kind of
\textit{backpropagation} feedback, a common type of
supervised learning\footnote{%
Note that we could also use a reward--based, {reinforcement
learning} rule (Suttton and Barto, 1998), in which system learns
its {optimal policy}:
\begin{eqnarray*}
innovation(t)=|reward(t)-penalty(t)|.
\end{eqnarray*}%
}
\begin{equation}
w_s(t+1)\,=\,w_s(t)-\eta \nabla J(t), \label{gradient}
\end{equation}
where $\eta $ is a small constant, called the \textit{step size},
or the \textit{learning rate,} and $\nabla J(n)$ denotes the
gradient of the `performance hyper--surface' at the $t$-th
iteration.

Now, the basic question about our local decision making process,
occurring under uncertainty at the intention formation faze
$\mathcal{F}$, is: Which alternative to choose? In our
path--integral language this reads: Which path (alternative)
should be given the highest probability weight $w$? This problem
can be either iteratively solved by the learning process
(\ref{gradient}), controlled by the $MONITOR$ feedback, which we
term \textit{algorithmic approach}, or by the local decision
making process under uncertainty, which we term \textit{heuristic
approach} (Ivancevic and Aidman, 2007). This qualitative analysis
is based on the micro--level interpretation of the Newtonian--like
action $S[x]$, given by (\ref{actmot}) and figuring both
processing speed $\dot{x}$ and LTM (i.e., the force--field
$\varphi(x)$, see next subsection). Here we consider three
different cases:\begin{enumerate}
    \item If the potential $\varphi(x)$ is not very dependent upon
    position $x(t)$, then the more direct paths contribute the
    most, as longer paths, with higher mean square velocities
    $[\dot{x}(t)]^2$ make the exponent more negative (after Wick rotation
    (\ref{Wick})).

    \item On the other hand, suppose that $\varphi(x)$ does indeed
    depend on position $x$. For simplicity, let the potential
    increase for the larger values of $x$. Then a direct path does
    not necessarily give the largest contribution to the overall
    transition probability, because the integrated value of the
    potential is higher than over another paths.

    \item Finally, consider a path that deviates wide-ly from the
    direct path. Then $\varphi(x)$ decreases over that path, but at the
    same time the velocity $\dot{x}$ increases. In this case, we
    expect that the increased velocity $\dot{x}$ would more than
    compensate for the decreased potential over the path.
\end{enumerate}
Therefore, the most important path (i.e., the path with the
highest weight $w$) would be the one for which any smaller
integrated value of the surrounding field potential $\varphi(x)$
is more than compensated for by an increase in kinetic--like
energy ${m\over2}\dot{x}^2$. In principle, this is neither the
most direct path, nor the longest path, but rather a middle way
between the two. Formally, it is the path along which the average
Lagrangian is minimal,
\begin{equation}
<{m\over2}\dot{x}^2+\varphi(x)>~\longrightarrow~\min, \label{DM}
\end{equation}
i.e., the \textit{path that requires minimal memory} (both LTM and
WM) and \textit{processing speed}. This mechanical result is
consistent with the `cognitive filter theory' of \textit{selective
attention} (Broadbent, 1958), which postulates a low level filter
that allows only a limited number of percepts to reach the brain
at any time. In this theory, the importance of conscious, directed
attention is minimized. The type of attention involving low level
filtering corresponds to the concept of \textit{early selection}.

Although we termed this `heuristic approach' in the sense that we
can instantly feel both the processing speed $\dot{x}$ and the LTM
field $\varphi(x)$ involved, there is clearly a psycho--physical
rule in the background, namely the averaging minimum relation
(\ref{DM}).

From the decision making point of view, all possible paths
(alternatives) represent the \textit{consequences} of decision
making. They are, by default, \textit{short--term consequences},
as they are modelled in the micro--time--level. However, the path
integral formalism allows calculation of the \textit{long--term
consequences}, just by extending the integration time,
$t_{fin}\to\infty$. Besides, this \textit{averaging decision
mechanics} -- choosing the optimal path -- actually performs the
`averaging lift' in the LSF: from the micro--level to the
macro--level.

For example, one of the simplest types of performance--degrading
disturbances in the LSF is what we term motivational fatigue -- a
motivational drag factor that slows the actors' prog-ress towards
their goal. There are two fundamentally different sources of this
motivational drag, both leading to apparently the same reduction
in performance: (a) tiredness / exhaustion and (b) satiation
(e.g., boredom). Both involve the same underlying mechanism (the
raising valence of the alternatives to continuing the action) but
the alternatives will differ considerably, depending on the
properties of the task, from self--preservation / recuperation in
the exhaustion case through to competing goals in the satiation
case.

The spatial representation of this motivational drag is relatively
simple: uni--dimensional LSF--coordinates may be sufficient for
most purposes, which makes it attractive for the initial
validation of our predictive model. Similarly uncomplicated
spatial representations can be achieved for what we term
motivational boost derived from the proximity to the goal
(including the well--known phenomenon of `the home stretch'): the
closer the goal (e.g., a finishing line) is perceived to be, the
stronger its `pulling power' (Lewin 1951, 1997). Combinations of
motivational drag and motivational boost effects may be of
particular interest in a range of applications. These combinations
can be modelled within relatively simple uni--dimensional
LSF--coordinate systems.

\section{Geometric Chaos and Topological Phase Transitions}

In this section we extend the LSF--formalism to incorporate
geometrical chaos (Ivancevic et al, 2008; Ivancevic, 2006c, 2008a)
and associated topological phase transitions.

It is well--known that on the basis of the {ergodic hypothesis},
statistical mechanics describes the physics of many--degrees of
freedom systems by replacing \emph{time averages} of the relevant
observables with \emph{ensemble averages}. Therefore, instead of
using statistical ensembles, we can investigate the Hamiltonian
(microscopic) dynamics of a system undergoing a phase transition.
The reason for tackling dynamics is twofold. First, there are
observables, like Lyapunov exponents, that are intrinsically
dynamical. Second, the geometrization of Hamiltonian dynamics in
terms of Riemannian geometry provides new observables and, in
general, an interesting framework to investigate the phenomenon of
phase transitions (Caiani et al, 1997; Pettini, 2007). The
geometrical formulation of the \emph{dynamics of conservative
systems} (see Ivancevic, 2006b, 2008a) was first used by Krylov
(1979) in his studies on the dynamical foundations of statistical
mechanics and subsequently became a standard tool to study
abstract systems in ergodic theory.

The simplest, mechanical--like LSF--action in the individual's
LSF--manifold $\bf\Sigma$ has a Riemannian locomotion form
(summation convention is always assumed)
\begin{equation}
S[q]={1\over2}\int_{t_{ini}}^{t_{fin}}[a_{ij}
\,\dot{q}^{i}\dot{q}^{j}-V(q)]\,dt, \label{locAct}
\end{equation}
where $a_{ij}$ is the `material' metric tensor that generates the
total `kinetic energy' of {cognitive} (loco)motions defined by
their configuration coordinates $q^i$ and velocities
$\dot{q}^{i}$, with the {motivational potential energy} $V(q)$ and
the standard Hamiltonian
\begin{equation}
H(p,q)=\sum_{i=1}^{N}\frac{1}{2}p_{i}^{2}+V(q), \label{Ham}
\end{equation}
where $p_{i}$ are the canonical (loco)motion momenta.

Dynamics of $N$ DOF mechanical--like systems with action
(\ref{locAct}) and Hamiltonian (\ref{Ham}) are commonly given by
the set of \emph{geodesic equations} (Ivancevic, 2006b, 2007a)
\begin{equation}
\frac{d^{2}q^{i}}{ds^{2}}+\Gamma _{jk}^{i}\frac{dq^{j}}{ds}\frac{dq^{k}}{ds}%
=0,  \label{geod-mot}
\end{equation}%
where $\Gamma _{jk}^{i}$ are the Christoffel symbols of the affine
Levi--Civita connection of the Riemannian LSF--manifold
$\bf\Sigma$.

Alternatively, a description of the extrema of the Hamilton's
action (\ref{locAct}) can be obtained using the \emph{Eisenhart
metric} (see Eisenhart, 1929) on an enlarged LSF space-time
manifold (given by $\{q^{0}\equiv t$, $q^{1},\ldots ,q^{N}\}$ plus
one real coordinate $q^{N+1}$), whose arc--length is
\begin{equation}
ds^{2}=-2V(q)(dq^{0})^{2}+a_{ij}dq^{i}dq^{j}+2dq^{0}dq^{N+1}.
\label{ds2E}
\end{equation}%
The manifold has a \emph{Lorentzian structure} (Pettini, 2007) and
the dynamical trajectories are those geode-sics satisfying the
condition $ds^{2}=Cdt^{2}$, where $C$ is a positive constant. In
this geometrical framework, the instability of the trajectories is
the instability of the geodesics, and it is completely determined
by the curvature properties of the LSF--manifold $\bf\Sigma$
according to the \emph{Jacobi equation} of geodesic deviation (see
Ivancevic, 2006b, 2007a)
\begin{equation}
\frac{D^{2}J^{i}}{ds^{2}}+R_{~jkm}^{i}\frac{dq^{j}}{ds}J^{k}\frac{dq^{m}}{ds}%
=0,  \label{eqJ}
\end{equation}%
whose solution $J$, usually called \emph{Jacobi variation field},
locally measures the distance between nearby geodesics; $D/ds$
stands for the {covariant derivative} along a geodesic and
$R_{~jkm}^{i}$ are the components of the {Riemann curvature
tensor} of the LSF--manifold $\bf\Sigma$.

Using the Eisenhart metric (\ref{ds2E}), the relevant part of the
Jacobi equation (\ref{eqJ}) is given by the \emph{tangent dynamics
equation} (Casetti et al, 1996; Caiani et al, 1997)
\begin{equation}
\frac{d^{2}J^{i}}{dt^{2}}+R_{~0k0}^{i}J^{k}=0,\qquad (i=1,\dots
,N), \label{eqdintang}
\end{equation}%
where the only non-vanishing components of the curvature tensor of
the LSF--manifold $\bf\Sigma$ are
\begin{eqnarray*}
R_{~0k0}^{i}=\partial ^{2}V/\partial q^{i}\partial q^{j}.
\end{eqnarray*}

The tangent dynamics equation (\ref{eqdintang}) is commonly used
to define \emph{Lyapunov exponents} in dynamical systems given by
the Riemannian action (\ref{locAct}) and Hamiltonian (\ref{Ham}),
using the formula (Casetti et al, 2000)
\begin{eqnarray}
\lambda _{1}&=&\lim_{t\rightarrow \infty }1/2t\log (\Sigma
_{i=1}^{N}[J_{i}^{2}(t) \label{Lyap1}\\
&+&J_{i}^{2}(t)]/\Sigma _{i=1}^{N}[J_{i}^{2}(0)+J_{i}^{2}(0)]).
\notag
\end{eqnarray}
Lyapunov exponents measure the \emph{strength of dynamical chaos}.

Now, to relate these results to topological phase transitions
within the LSF--manifold $\bf\Sigma$, recall that any two
high--dimensional manifolds $\Sigma _{v}$ and $\Sigma _{v^{\prime
}}$ have the same topology if they can be continuously and
differentiably deformed into one another, that is if they are
diffeomorphic. Thus by \emph{topology change} the `loss of
diffeomorphicity' is meant (Pettini, 2007). In this respect, the
so--called \emph{topological theorem} (Franzosi and Pettinni,
2004) says that non--analyti-city is the `shadow' of a more
fundamental phenomenon occurring in the system's configuration
manifold (in our case the LSF--manifold): a topology change within
the family of equipotential hypersurfaces
\begin{eqnarray*}
\Sigma _{v}=\{(q^{1},\dots ,q^{N})\in \mathbb{R}^{N}|\
V(q^{1},\dots ,q^{N})=v\},
\end{eqnarray*}%
where $V$ and $q^{i}$ are the microscopic interaction potential
and coordinates respectively. This topological approach to PTs
stems from the numerical study of the dynamical counterpart of
phase transitions, and precisely from the observation of
discontinuous or cuspy patterns displayed by the largest Lyapunov
exponent $\lambda _{1}$ at the \emph{transition energy} (Casetti
et al, 2000). Lyapunov exponents cannot be measured in laboratory
experiments, at variance with thermodynamic observables, thus,
being genuine dynamical observables they are only be estimated in
numerical
simulations of the microscopic dynamics. If there are critical points of $V$ in configuration space, that is points $%
q_{c}=[{\overline{q}}_{1},\dots ,{\overline{q}}_{N}]$ such that
$\left. \nabla V(q)\right\vert _{q=q_{c}}=0$, according to the
\emph{Morse Lemma} (see e.g., Hirsch, 1976), in the neighborhood
of any {critical point} $q_{c}$
there always exists a coordinate system\\ ${\tilde{q}}(t)=[{\tilde{q}}%
^{1}(t),..,{\tilde{q}}^{N}(t)]$ for which
\begin{equation}
V({\tilde{q}})=V(q_{c})-{\tilde{q}}_{1}^{2}-\dots -{\tilde{q}}_{k}^{2}+{%
\tilde{q}}_{k+1}^{2}+\dots +{\tilde{q}}_{N}^{2},
\label{morsechart}
\end{equation}%
where $k$ is the {index of the critical point}, i.e., the number
of negative eigenvalues of the Hessian of the potential energy
$V$. In the neighborhood of a critical point of the LSF--manifold
$\bf\Sigma$, (\ref{morsechart}) yields
\begin{eqnarray*}
\partial ^{2}V/\partial q^{i}\partial q^{j}=\pm \delta _{ij},
\end{eqnarray*}%
which gives $k$ unstable directions which contribute to the
exponential growth of the norm of the tangent vector $J$ (Casetti
et all, 2000).

This means that the strength of dynamical chaos within the
individual's LSF--manifold $\bf\Sigma$, measured by the largest
Lyapunov exponent $\lambda _{1}$ given by (\ref{Lyap1}), is
affected by the existence of critical points $q_{c}$ of the
potential energy $V(q)$. However, as $V(q)$ is bounded below, it
is a good {Morse function}, with no vanishing eigenvalues of its
Hessian matrix. According to \emph{Morse theory},
 the existence of critical points of $V$ is
associated with topology changes of the hypersurfaces $\{\Sigma
_{v}\}_{v\in \mathbb{R}}$.

More precisely, let $V_{N}(q_{1},\dots ,q_{N}):R^{N}\rightarrow
R$, be a
smooth, bounded from below, finite-range and confining potential\footnote{%
These requirements for $V$ are fulfilled by standard interatomic
and intermolecular interaction potentials, as well as by classical
spin potentials.}. Denote by $\Sigma _{v}=V^{-1}(v)$, $v\in R$,
its level sets,
or equipotential hypersurfaces, in the LSF--manifold $\bf\Sigma$. Then let $\bar{v}%
=v/N $ be the potential energy per degree of freedom. If there exists $N_{0}$%
, and if for any pair of values $\bar{v}$ and $\bar{v}^{\prime }$
belonging
to a given interval $I_{\bar{v}}=[\bar{v}_{0},\bar{v}_{1}]$ and for any $%
N>N_{0} $ then the sequence of the Helmoltz free energies\\
$\{F_{N}(\beta
)\}_{N\in \mathbb{N}}$ -- where $\beta =1/T$ ($T$ is the temperature) and $%
\beta \in I_{\beta }=(\beta (\bar{v}_{0}),\beta (\bar{v}_{1}))$ --
is uniformly convergent at least in $C^{2}(I_{\beta })$ [the space
of twice
differentiable functions in the interval $I_{\beta }$], so that $%
\lim_{N\rightarrow \infty }F_{N}\in C^{2}(I_{\beta })$ and neither
first nor second order phase transitions can occur in the
(inverse) temperature interval $(\beta (\bar{v}_{0}),\beta
(\bar{v}_{1}))$, where the inverse temperature is defined as
(Pettini, 2007)
\begin{eqnarray*}
&&\beta (\bar{v})=\partial S_{N}^{(-)}(\bar{v})/\partial \bar{v},\qquad \text{%
while}\\ &&S_{N}^{(-)}(\bar{v})=N^{-1}\log \int_{V(q)\leq
\bar{v}N}\ d^{N}q
\end{eqnarray*}%
is one of the possible definitions of the microcanonical
configurational entropy. The intensive variable $\bar{v}$ has been
introduced to ease the comparison between quantities computed at
different $N$-values.

This theorem means that a topology change of the $\{\Sigma
_{v}\}_{v\in \mathbb{R}}$ at some $v_{c}$ is a {necessary}
condition for a phase transition to take place at the
corresponding energy value. The topology changes implied here are
those described within the framework of Morse theory through
`attachment of handles' to the LSF--manifold $\bf\Sigma$ (Hirsch,
1976).

In the LSF path--integral language, we can say that suitable
topology changes of equipotential submanifolds of the individual's
LSF--manifold $\bf\Sigma$ can entail thermodynamic--like phase
transitions, according to the general formula:
\begin{eqnarray*}
&&\langle {\rm phase~out}\,|\,{\rm phase~in}\rangle\,:=\\
&&\put(0,0){\LARGE $\int$}_{\rm
~~topology-change}\put(-650,20){\small
$\bf\Sigma$}\mathcal{D}[w\Phi]\, {\mathrm e}^{\mathrm i S[\Phi]}.
\end{eqnarray*}
The statistical behavior of the LSF--(loco)moti-on system
(\ref{locAct}) with the standard Hamiltonian (\ref{Ham}) is
encompassed, in the canonical ensemble, by its {partition
function}, given by the phase--space path integral (Ivancevic,
2007a, 2008a)
\begin{equation}
Z_{N}=\put(0,0){\LARGE $\int$}_{\rm
~~top-ch}\put(-310,20){\small$\bf\Sigma$}\mathcal{D}[p]\mathcal{D}[q]\exp \{ \mathrm{i}\int_{t}^{t{%
^{\prime }}}[p\dot{q}-H(p,q)]\,d\tau \} , \label{PI1}
\end{equation}%
where we have used the shorthand notation
\begin{eqnarray*}
\put(0,0){\LARGE $\int$}_{\rm
~~top-ch}\put(-310,20){\small$\bf\Sigma$}\mathcal{D}[p]\mathcal{D}[q]\equiv
\int \prod_{\tau }\frac{dq(\tau )dp(\tau )}{2\pi }.
\end{eqnarray*}%
The phase--space path integral (\ref{PI1}) can be calculated as
the partition function (Franzosi et al, 2000)
\begin{eqnarray}
&&Z_{N}(\beta ) =\int \prod_{i=1}^{N}dp_{i}dq^{i}{\rm e}^{-\beta
H(p,q)} \notag\\ &&=\left( \frac{\pi }{\beta }\right)
^{\frac{N}{2}}\!\!\int
\prod_{i=1}^{N}dq^{i}{\rm e}^{-\beta V(q)}  \notag \\
&&=\left( \frac{\pi }{\beta }\right)
^{\frac{N}{2}}\int_{0}^{\infty }dv\,{\rm e}^{-\beta v}\int_{\Sigma
_{v}}\frac{d\sigma }{\Vert \nabla V\Vert }\!\!,, \label{zeta}
\end{eqnarray}%
where the last term is written using the so--called \emph{co--area
formula} (Federer, 1969), and $v$ labels the equipotential
hypersurfaces $\Sigma _{v}$ of the LSF--manifold $\bf\Sigma$,
\begin{eqnarray*}
\Sigma _{v}=\{(q^{1},\dots ,q^{N})\in \mathbb{R}^{N}|V(q^{1},\dots
,q^{N})=v\}.
\end{eqnarray*}%
Equation (\ref{zeta}) shows that the relevant statistical
information is contained in the canonical configurational
partition function
\begin{eqnarray*}
Z_{N}^{C}=\int \prod dq^{i}V(q){\rm e}^{-\beta V(q)}.
\end{eqnarray*}%
Note that $Z_{N}^{C}$ is decomposed, in the last term of
(\ref{zeta}), into an infinite summation of geometric integrals,
\begin{eqnarray*}
\int_{\Sigma _{v}}d\sigma \,/\Vert \nabla V\Vert ,
\end{eqnarray*}%
defined on the $\{\Sigma _{v}\}_{v\in \mathbb{R}}$. Once the
microscopic interaction potential $V(q)$ is given, the
configuration space of the system is automatically foliated into
the family $\{\Sigma _{v}\}_{v\in \mathbb{R}}$ of these
equipotential hypersurfaces. Now, from standard statistical
mechanical arguments we know that, at any given value of the
inverse
temperature $\beta $, the larger the number $N$, the closer to $%
\Sigma _{v}\equiv \Sigma _{u_{\beta }}$ are the microstates that
significantly contribute to the averages, computed through
$Z_{N}(\beta )$, of thermodynamic observables. The hypersurface
$\Sigma _{u_{\beta }}$ is the one associated with
\begin{eqnarray*}
u_{\beta }=(Z_{N}^{C})^{-1}\int \prod dq^{i}V(q){\rm e}^{-\beta
V(q)},
\end{eqnarray*}%
the average potential energy computed at a given $\beta $. Thus, at any $%
\beta $, if $N$ is very large the effective support of the
canonical measure shrinks very close to a single $\Sigma
_{v}=\Sigma _{u_{\beta }}$. Hence, the basic origin of a phase
transition lies in a suitable topology change of the $\{\Sigma
_{v}\}$, occurring at some $v_{c}$ (Franzosi et al, 2000). This
topology change induces the singular behavior of the thermodynamic
observables at a phase transition. It is conjectured that the
counterpart of a phase transition is a breaking of
diffeomorphicity among the surfaces $\Sigma _{v}$, it is
appropriate to choose a {diffeomorphism invariant} to probe if and
how the topology of the $\Sigma _{v}$ changes as a function of
$v$. Fortunately, such a topological invariant exists, the {Euler
characteristic} of the LSF--manifold $\bf\Sigma$, defined by
(Ivancevic, 2007a)
\begin{equation}
\chi (\Sigma )=\sum_{k=0}^{N}(-1)^{k}b_{k}(\Sigma ),  \label{chi}
\end{equation}%
where the {Betti numbers} $b_{k}(\Sigma )$ are diffeomorphism invariants.\footnote{%
The Betti numbers $b_{k}$ are the dimensions of the de Rham's
cohomology vector spaces $H^{k}(\Sigma ;\mathbb{R})$ (therefore
the $b_{k}$ are integers).} This homological formula can be
simplified by the use of the {Gauss--Bonnet--Hopf theorem}, that
relates $\chi (\Sigma )$ with the total {Gauss--Kronecker
curvature} $K_{G}$ of the LSF--manifold $\bf\Sigma$
\begin{equation}
\chi (\Sigma )=\int_{\Sigma }K_{G}\,d\sigma, \label{gaussbonnet}
\end{equation}%
where
\begin{eqnarray*}
d\sigma =\sqrt{det(a)}dx^{1}dx^{2}\cdots dx^{n}
\end{eqnarray*}%
is the invariant volume measure of the LSF--manifold $\bf\Sigma$
and $a$ is the determinant of the LSF metric tensor $a_{ij}$
(Ivancevic, 2008a).

The domain of validity of the `quantum' is not restricted to the
microscopic world (Umezawa, 1993). There are macroscopic features
of classically behaving systems, which cannot be explained without
recourse to the quantum dynamics. This field theoretic model leads
to the view of the phase transition as a condensation that is
comparable to the formation of fog and rain drops from water
vapor, and that might serve to model both the gamma and beta phase
transitions. According to such a model, the production of activity
with long-range correlation in the brain takes place through the
mechanism of spontaneous breakdown of symmetry (SBS), which has
for decades been shown to describe long-range correlation in
condensed matter physics. The adoption of such a field theoretic
approach enables modelling of the whole cerebral hemisphere and
its hierarchy of components down to the atomic level as a fully
integrated macroscopic quantum system, namely as a macroscopic
system which is a quantum system not in the trivial sense that it
is made, like all existing matter, by quantum components such as
atoms and molecules, but in the sense that some of its macroscopic
properties can best be described with recourse to quantum dynamics
(see Freeman and Vitiello, 2006 and references therein).

Phase transitions can also be associated with autonomous robot
competence levels, as informal specifications of desired classes
of behaviors for robots over all environments they will encounter,
as described by Brooks' subsumption architecture approach. The
distributed network of augmented finite--state machines can exist
in different phases or modalities of their state--space variables,
which determine the systems intrinsic behavior. The phase
transition represented by this approach is triggered by either
internal (a set--point) or external (a command) control stimuli,
such as a command to transition from a sleep mode to awake mode,
or walking to running.

\section{Joint Action of Several\\ Agents}

In this section we propose an LSF--based model of the joint action
of two or more actors, where actors can be both humans and robots.
This joint action takes place in the joint LSF manifold
$\Sigma_J$, composed of individual LSF manifolds $\Sigma_\alpha,
\Sigma_\beta,...$. It has a sophisticated geometrical and
dynamical structure as follows.

To model the dynamics of the two--actor co--action, we propose to
associate each of the actors with a set of their own time
dependent trajectories, which constitutes an $n-$dimensional
Riemannian LSF--manifold, $\Sigma_{\alpha }=\{x^{i}(t_{i})\}$ and
$\Sigma_{\beta }=\{y^{j}(t_{j})\}$, respectively. Their associated
tangent bundles contain their individual $n$D (loco)motion
velocities, $T\Sigma_{\alpha }=\{\dot{x}^{i}(t_{i})=dx^{i}/dt_{i}\}$ and $%
T\Sigma_{\beta }=\{\dot{y}^{j}(t_{j})=dy^{j}/dt_{j}\}.$ Further,
following the general LSF--formalism, outlined in the
introduction, we use the modelling machinery consisting of: (i)
Adaptive joint action at the top--master level, describing the
externally--appearing deterministic, continuous and smooth
dynamics, and (ii) Corresponding adaptive path integral
(\ref{pathInt}) at the bottom--slave level, describing a wildly
fluctuating dynamics including both continuous trajectories and
Markov chains. This lower--level joint dynamics can be further
discretized into a partition function of the corresponding
statistical dynamics.

The smooth joint action with two terms, representing
cognitive/motivational potential energy and physical kinetic
energy, is formally given by:
\begin{eqnarray}
&&A[x,y;t_{i},t_{j}] =  \notag \\
&&\frac{1}{2}\int_{t_{i}}\int_{t_{j}}\alpha _{i}\beta _{j}\,\delta
(I_{ij}^{2})\,\,\dot{x}^{i}(t_{i})\,\dot{y}^{j}(t_{j})\,\,dt_{i}dt_{j}
\notag \\
&&+{\frac{1}{2}}\int_{t}g_{ij}\,\dot{x}^{i}(t)\dot{x}^{j}(t)\,dt,  \label{Fey1} \\
&&\text{with\qquad }I_{ij}^{2} =\left[
x^{i}(t_{i})-y^{j}(t_{j})\right] ^{2},
\notag \\
&&\text{where \ \ }IN \leq t_{i},t_{j},t\leq OUT.\hspace{2cm}
\notag
\end{eqnarray}
The first term in (\ref{Fey1}) represents {potential energy of the
cognitive/motivational interaction} between the two agents $\alpha
_{i}$ and
$\beta _{j}$.\footnote{%
Although, formally, this term contains cognitive velocities, it
still represents `potential energy' from the physical point of
view.} It is a double integral over a delta function of the square
of interval $I^{2}$ between two points on the paths in their
Life--Spaces; thus, interaction occurs only when this interval,
representing the motivational cognitive distance between the two
agents, vanishes. Note that the cognitive (loco) motions of the
two agents $\alpha _{i}[x^{i}(t_{i})]$ and $\beta
_{j}[y^{j}(t_{j})]$, generally occur at different times $t_{i}$
and $t_{j}$\ unless $t_{i}=t_{j},$ when {cognitive
synchronization} occurs.

The second term in (\ref{Fey1}) represents {kinetic energy of the
physical interaction}. Namely, when the cognitive synchronization
in the first term takes place, the second term of physical kinetic
energy is activated in the common manifold, which is one of the
agents' Life Spaces, say $\Sigma_{\alpha }=\{x^{i}(t_{i})\}$.

Conversely, if we have a need to represent coaction of three actors, say $\alpha _{i}$, $%
\beta _{j}$ and $\gamma _{k}$ (e.g., $\alpha _{i}$ in charge of
acceleration, $\beta _{j}$ -- deceleration and $\gamma _{k}-$
steering), we can associate each of them with an $n$D Riemannian
Life--Space manifold,
$\Sigma_{\alpha }=\{x^{i}(t_{i})\}$, $\Sigma_{\beta }=\{y^{j}(t_{j})\}$, and $%
\Sigma_{\gamma }=\{z^{k}(t_{k})\},$ respectively, with the
corresponding tangent
bundles containing their individual (loco) motion velocities, $T\Sigma_{\alpha }=\{%
\dot{x}^{i}(t_{i})=dx^{i}/dt_{i}\}$, $T\Sigma_{\beta }=\{\dot{y}%
^{j}(t_{j})=dy^{j}/dt_{j}\}$ and $T\Sigma_\gamma =\{\dot{z}^{k}(t_{k})$ $=dz^{k}/dt_{k}%
\}.$ Then, instead of (\ref{Fey1}) we have
\begin{eqnarray}
&&A[t_{i},t_{j},t_{k};t] =  \notag  \label{Fey11} \\
&&\frac{1}{2}\int_{t_{i}}\int_{t_{j}}\int_{t_{k}}\alpha
_{i}(t_{i})\beta
_{j}\,(t_{j})\,\gamma _{k}\,(t_{k})\times   \notag \\
&&\delta (I_{ijk}^{2})\,\,\dot{x}^{i}(t_{i})\,\dot{y}^{j}(t_{j})\,\dot{z}%
^{k}(t_{k})\,dt_{i}dt_{j}dt_{k}  \notag \\
&&+~{\frac{1}{2}}\int_{t}W_{rs}^{M}(t,q,\dot{q})\,\dot{q}^{r}\dot{q}%
^{s}\,dt,\,\qquad   \label{Fey2}
\end{eqnarray}%
\begin{eqnarray*}
&&\text{where }IN \leq t_{i},t_{j},t_{k},t\leq OUT,\qquad \text{with} \\
&&I_{ijk}^{2}
=[x^{i}(t_{i})-y^{j}(t_{j})]^{2}+[y^{j}(t_{j})-z^{k}(t_{k})]^{2} \\
&&+~[z^{k}(t_{k})-x^{i}(t_{i})]^{2},
\end{eqnarray*}

Due to an {intrinsic chaotic coupling}, the three--actor (or,
$n-$actor, $n>3$) joint action (\ref{Fey2}) has a considerably
more complicated geometrical structure then the bilateral
co--action (\ref{Fey1}).\footnote{Recall that the {necessary}
condition for chaos in continuous temporal or spatio-temporal
systems is to have
three variables with nonlinear couplings between them.} It actually happens in the common $3n$D {%
Finsler manifold} $\Sigma_{J}=\Sigma_{\alpha }\cup \Sigma_{\beta
}\cup \Sigma_\gamma$,
parameterized by the local joint coordinates dependent on the common time $t$%
. That is, $\Sigma_{J}=\{q^{r}(t),\,r=1,...,3n\}.$ Geometry of the
joint manifold
$\Sigma_{J}$ is defined by the {Finsler metric function} $%
ds=F(q^{r},dq^{r}),$ defined by
\begin{equation}
F^{2}(q,\dot{q})=g_{rs}(q,\dot{q})\dot{q}^{r}\dot{q}^{s},
\label{Fins1}
\end{equation}%
and the \emph{Finsler tensor} $C_{rst}(q,\dot{q}),$ defined by
(Ivancevic, 2007a)
\begin{equation}
C_{rst}(q,\dot{q})=\frac{1}{4}\frac{\partial
^{3}F^{2}(q,\dot{q})}{\partial
\dot{q}^{r}\partial \dot{q}^{s}\partial \dot{q}^{t}}=\frac{1}{2}\frac{%
\partial g_{rs}}{\partial \dot{q}^{r}\partial \dot{q}^{s}}.  \label{Fins2}
\end{equation}%
From the Finsler definitions (\ref{Fins1})--(\ref{Fins2}), it
follows that the partial interaction manifolds, $\Sigma_{\alpha
}\cup \Sigma_{\beta },$ $\Sigma_{\beta }\cup \Sigma_{y}$ and
$\Sigma_{\alpha }\cup \Sigma_{y}$, have Riemannian structures with
the
corresponding interaction kinetic energies,
\begin{eqnarray*}
&&T_{\alpha \beta }=\frac{1}{2}%
g_{ij}\dot{x}^{i}\dot{y}^{j},\qquad T_{\alpha \gamma }=\frac{1}{2}g_{ik}\dot{x}%
^{i}\dot{z}^{k},\\ &&T_{\beta \gamma }=\frac{1}{2}g_{jk}\dot{y}^{j}\dot{z}%
^{k}.
\end{eqnarray*}

At the slave LSF--level, the adaptive path integral, representing
an $\infty-$dimensional neural network, corresponding to the
adaptive bilateral joint action (\ref{Fey1}), reads
\begin{equation}
\langle OUT|IN\rangle :=\put(0,0){\LARGE $\int$}\put(15,15){\small
$\bf\Sigma$}\quad\,\mathcal{D}[w,x,y]\, {\mathrm e}^{\mathrm i
A[x,y;t_{i},t_{j}]}, \label{pathInt}
\end{equation}
where the Lebesgue integration is performed over all continuous paths $%
x^{i}=x^{i}(t_{i})$ and $y^{j}=y^{j}(t_{j})$, while summation is
performed over all associated discrete Markov fluctuations and\\
jumps. The symbolic
differential in the path integral (\ref{pathInt}) represents an {%
adaptive path measure}, defined as a weighted product
\begin{eqnarray}
\mathcal{D}[w,x,y]=\lim_{N\rightarrow \infty
}\prod_{s=1}^{N}w_{ij}^{s}dx^{i}dy^{j}, \label{prod}\\
({i,j=1,...,n}).\notag
\end{eqnarray}

Similarly, in case of the triple joint action, the adaptive path
integral reads,
\begin{equation}
\langle OUT|IN\rangle :=\put(0,0){\LARGE $\int$}\put(15,15){\small
$\bf\Sigma$}\quad\,\mathcal{D}[w;x,y,z;q]\, {\mathrm e}^{\mathrm i
A[t_{i},t_{j},t_{k};t]}, \label{pathInt2}
\end{equation}
with the adaptive path measure defined by%
\begin{eqnarray}
\mathcal{D}[w;x,y,z;q]=\lim_{N\rightarrow \infty
}\prod_{S=1}^{N}w_{ijkr}^{S}dx^{i}dy^{j}dz^{k}dq^{r},\notag \\
(i,j,k=1,...,n;~r=1,...,3n). \hspace{1cm} \label{prod2}
\end{eqnarray}

The proposed path integral approach represents a new family of
more expansive function-representation methods, which is now
capable of representing input/output behavior of more than one
actor. However, as we add the second and subsequent actors to the
model, the requirements for the rigorous geometrical
representations of their respective LSFs become nontrivial. For a
single actor or a two--actor co--action the Riemannian geometry
was sufficient, but it becomes insufficient for modelling the
$n$--actor (with $n\geq 3$) joint action, due to an {intrinsic
chaotic coupling} between the individual actors' LSFs. To model an
$n$--actor joint LSF, we have to use the Finsler geometry, which
is a generalization of the Riemannian one. This progression may
seem trivial, both from standard psychological point of view, and
from computational point of view, but it is not trivial from the
geometrical perspective.

Our extended LSF formalism is closely related to the
Haken-Kelso-Bunz (HKB) model of self-organization in the human
motor system (Haken et al, 1985; Kelso, 1995), including:
multi-stability, phase transitions and hysteresis effects,
presenting a contrary view to the purely feedback driven neural
systems. HKB uses the concepts of synergetics (order parameters,
control parameters, instability, etc) and the mathematical tools
of nonlinearly coupled (nonlinear) dynamical systems to account
for self-organized behavior both at the cooperative, coordinative
level and at the level of the individual coordinating elements.
The HKB model stands as a building block upon which numerous
extensions and elaborations have been constructed. In particular,
it has been possible to derive it from a realistic model of the
cortical sheet in which neural areas undergo a reorganization that
is mediated by intra- and inter-cortical connections. Also, the
HKB model describes phase transitions (`switches') in coordinated
human movement as follows: (i) when the agent begins in the
anti-phase mode and speed of movement is increased, a spontaneous
switch to symmetrical, in-phase movement occurs; (ii) this
transition happens swiftly at a certain critical frequency; (iii)
after the switch has occurred and the movement rate is now
decreased the subject remains in the symmetrical mode, i.e. she
does not switch back; and (iv) no such transitions occur if the
subject begins with symmetrical, in-phase movements. The HKB
dynamics of the order parameter relative phase as is given by a
nonlinear first-order ODE:
$$\dot{\phi} = (\alpha + 2 \beta r^2) \sin\phi - \beta r^2
\sin2\phi,
$$
where $\phi$ is the phase relation (that characterizes the
observed patterns of behavior, changes abruptly at the transition
and is only weakly dependent on parameters outside the phase
transition), $r$ is the oscillator amplitude, while $\alpha,\beta$
are coupling parameters (from which the critical frequency where
the phase transition occurs can be calculated).

From a quantum perspective, closely related to the LSF model are
the recent developments of Hong and Newell (2008a, 2008b) in motor
control that deal with feedback information and environmental
uncertainty. The probabilistic nature of human action can be
characterized by entropies at the level of the organism, task, and
environment. Systematic changes in motor adaptation are
characterized as task--organism and environment--organism
tradeoffs in entropy. Such compensatory adaptations lead to a view
of goal--directed motor control as the product of an underlying
conservation of entropy across the task-organism-environment
system. The conservation of entropy supports the view that context
dependent adaptations in human goal--directed action are guided
fundamentally by natural law and provides a novel means of
examining human motor behavior. This is fundamentally related to
the \emph{Heisenberg uncertainty principle} and further support
the argument for the primacy of a probabilistic approach toward
the study of bio-psychological systems. \onecolumn
\begin{figure}[ht]
\centerline{\includegraphics[width=16cm]{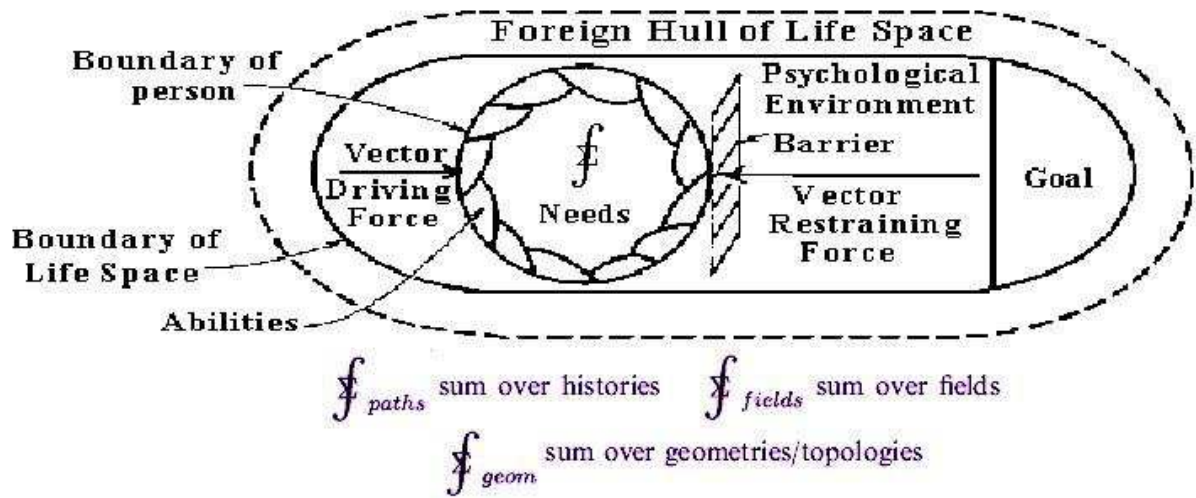}}
\caption{Diagram of the \textit{life space foam}: classical
representation of Lewinian life space, with an adaptive path
integral $\put(0,0){\LARGE $\int$}\put(10,15){\small
$\bf\Sigma$}\quad$ (denoting integration over continuous paths and
summation over discrete Markov jumps) acting inside it and
generating microscopic fluctuation dynamics.} \label{LifeSpace}
\end{figure}
\begin{figure}[htb]
\centerline{\includegraphics[height=6cm]{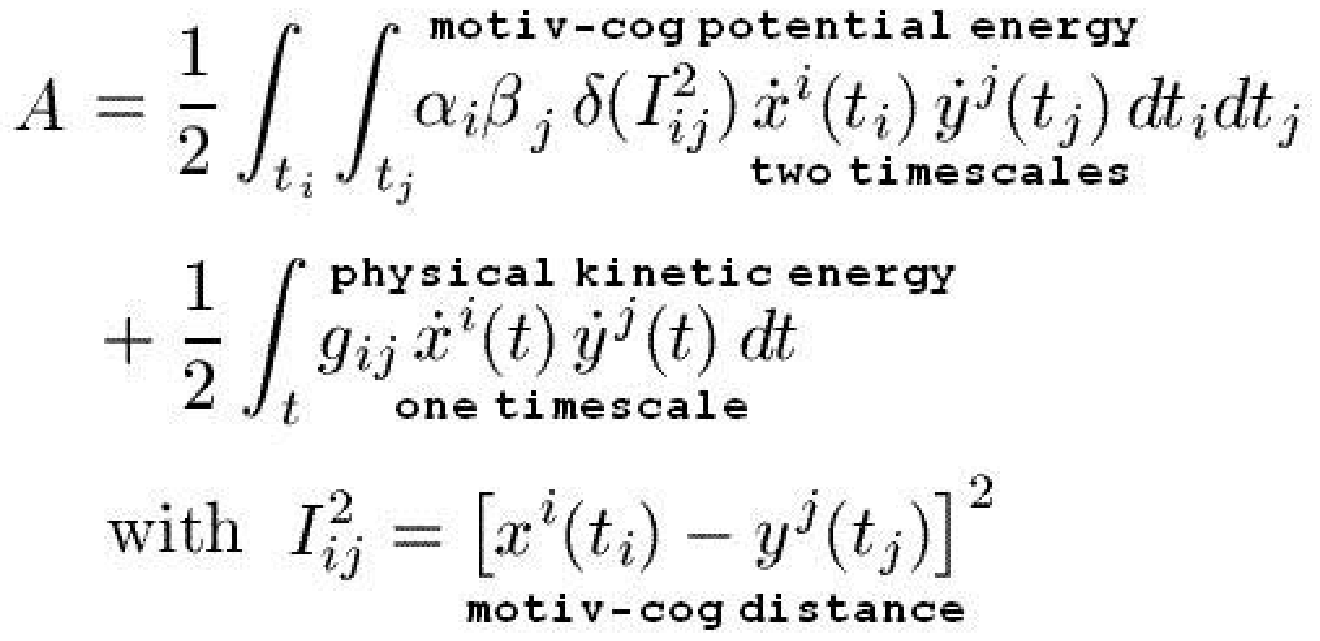}}
\caption{{\protect\small {Feynman action for modelling human joint
action, including potential energy (motivational cognition) in two
timescales, physical energy in a single timescale (after
synchronization has already occurred), and the distance between
two agents in the motivational cognition space.}}}
\label{IvJAction}
\end{figure}\twocolumn

\section{Conclusion}

General stochastic and quantum dynamics, developed in a framework
of Feynman path integrals, have recently been applied by Ivancevic
and Aidman (2007) to Lewinian field--theoretic psychodynamics,
resulting in the development of a new concept of \emph{Life--Space
Foam} (LSF) as a natural medium for motivational (MD) and
cognitive (CD) psychodynamics. According to the LSF--formalism,
the classic Lewinian life space can be macroscopically represented
as a smooth manifold with steady force--fields and behavioral
paths, while at the microscopic level it is more realistically
represented as a collection of wildly fluctuating force--fields,
(loco)mo-tion paths and local geometries (and topologies with
holes). This paper extends the LSF--model to incorporate the
notion of phase transitions and complements it with embedded
geometrical chaos. As a result, the extended LSF--model is able to
rigorously represent co--action by two or more human--like agents
in the common LSF--manifold. The extended LSF--model is also
related to the HKB--model of self-organi-zation in the human motor
system, presenting a contrary view to the purely feedback driven
neural systems, as well as Hong--Newell entropy--approach to
adaptation in human goal--directed motor control.\\

\bigbreak

\noindent{\bf References}\\

Amari S. Dynamics of pattern formation in lateral-inhibition type
neural fields. Biol Cybern 1977;{27}:77--87.

Bernstein NA, Latash ML and Turvey MT (Eds). Dexterity and its
development. Hillsdale NJ England: Lawrence Erlbaum Associates
1996.

Bernstein NA. Some emergent problems of the regulation of motor
acts. In: Whiting H (ed) Human Motor Actions: Bernstein Reassessed
343--358. North Holland, Amsterdam 1982.

Bernstein NA. The Coordination and Regulation of Movements.
Pergamon London 1967.

Broadbent DE. Perception and communications. Pergamon Press,
London 1958.

Brooks RA. A Robust Layered Control System for a Mobile Robot.
IEEE Trans Rob Aut 1986;(2)1:14--23.

Brooks RA. A robot that walks: Emergent behavior form a carefully
evolved network. Neu Comp 1989;12:253--262.

Brooks RA. Elephants Don't Play Chess. Rob Aut Sys 1990;{6}:3--15.

Busemeyer JR, Wang Z and Townsend JT. Quantum dynamics of human
decision-making. J Math Psych 2006;{50}:220--241.

Caiani L, Casetti L, Clementi C and Pettini M. Geometry of
Dynamics Lyapunov Exponents and Phase Transitions. Phys Rev Lett
1997;{79}:4361-4364.

Casetti L, Clementi C and Pettini M. Riemannian theory of
Hamiltonian chaos and Lyapunov exponents. Phys Rev E
1996;{54}:5969.

Casetti L, Pettini M and Cohen EGD. Geometric Approach to
Hamiltonian Dynamics and Statistical Mechanics. Phys Rep
2000;{337}:237-341.

Droll, JA, Hayhoe, MM, Triesch, J and Sullivan, BT. Task demands
control acquisition and storage of visual information. J Exp
Psych: Hum Perc Perf 2005;31:1416–1438.

Eisenhart LP. Dynamical trajectories and geodesics. Math Ann
1929;{30}:591--606.

Federer H. Geometric Measure Theory.\\ Springer New York 1969.

Fogassi L, Ferrari PF, Gesierich B, Rozzi S, Chersi F and
Rizzolatti G. Parietal lobe: From action organization to intention
understanding. Science 2005;{29}:662--667.

Franzosi R and Pettini M. Theorem on the origin of Phase
Transitions. Phys Rev Lett 2004;{92}:060601.

Franzosi R, Pettini M, Spinelli L. Topology and phase transitions:
a paradigmatic evidence. Phys Rev Lett 2000;{84}:2774--2777.

Freeman WJ and Vitiello G. Nonlinear brain dynamics as macroscopic
manifestation of underlying many-body field dynamics. Phys Life
Rev 2006;(3)2:93--118.

Freeman WJ. Mass Action in the Nervous System. Acad Press, New
York 1975/2004.

Freeman WJ. Neurodynamics. An Exploration of Mesoscopic Brain
Dynamics. Springer, London 2000.

Gardiner C.W. Handbook of Stochastic Methods for Physics Chemistry
and Natural Sciences 2nd ed. Springer, New York 1985.

Glimcher PW. Indeterminacy in brain and behaviour. Ann Rev Psych
2005;{56}:25--56.

Gold M. A Kurt Lewin Reader the Complete Social Scientist. Am
Psych Assoc Washington 1999.

Haken H. Advanced Synergetics: Instability Hierarchies of
Self-Organizing Systems and Devices (3nd ed). Springer, Berlin
1993.

Haken H. Principles of Brain Functioning: A Synergetic Approach to
Brain Activity Behavior and Cognition. Springer, Berlin 1996.

Haken H. Synergetics: An Introduction (3rd ed). Springer, Berlin
1983.

Haken, H, Kelso, JAS and Bunz H. A theoretical model of phase
transitions in human hand movements. Biol Cybern 1985;51:347--356.

Heermann DW. Computer Simulation Methods in Theoretical Physics
(2nd ed). Springer, Berlin 1990.

Hirsch MW. Differential Topology. Springer, New York 1976.

Hong SL, Newell KM. Entropy compensation in human motor
adaptation. Chaos\\ 2008b;18(1):013108.

Hong SL, Newell KM. Entropy conservation in the control of human
action. Nonl Dyn Psych Life Sci 2008a;12(2):163-190.

Ivancevic T, Jain L, Pattison J and Hariz A. Nonlinear Dynamics
and Chaos Methods in Neurodynamics and Complex Data Analysis. Nonl
Dyn 2008(in press, on line first, Springer).

Ivancevic V and Aidman E. Life-space foam: A medium for
motivational and cognitive dynamics. Physica A
2007;{382}:616--630.

Ivancevic V and Ivancevic T. Applied Differential Geometry: A
Modern Introduction. World Scientific, Singapore 2007a.

Ivancevic V and Ivancevic T. Complex Nonlinearity: Chaos, Phase
Transitions, Topology Change and Path Integrals. Springer, Berlin
2008a.

Ivancevic V and Ivancevic T. Computational Mind: A Complex
Dynamics Perspective.\\ Springer, Berlin 2007c.

Ivancevic V and Ivancevic T. Geometrical Dynamics of Complex
Systems: A Unified Modelling Approach to Physics Control
Biomechanics Neurodynamics and Psycho-Socio-Economical Dynamics.
Springer, Dordrecht 2006b.

Ivancevic V and Ivancevic T. High--Dimen-sional Chaotic and
Attractor Systems. Springer, Berlin 2006c.

Ivancevic V and Ivancevic T. Natural Biodynamics. World
Scientific, Singapore 2006a.

Ivancevic V and Ivancevic T. Neuro--Fuzzy Associative Machinery
for Comprehensive Brain and Cognition Modelling. Springer, Berlin
2007b.

Ivancevic V and Ivancevic T. Quantum Leap: From Dirac and Feynman
Across the Universe to Human Body and Mind. World Scientific,
Singapore 2008b.

Izhikevich EM and Edelman GM. Large-Scale Model of Mammalian
Thalamocortical Systems. PNAS 2008;{105}:3593--3598.

Kelso, JAS. Dynamic Patterns: The Self Organization of Brain and
Behavior. MIT Press, Cambridge 1995.

Knill DC, Maloney LT and Trommershauser J. Sensorimotor Processing
and Goal-Directed Movement. J Vis 2007;7(5)i:1-2.

Knoblich G and Jordan S. Action coordination in individuals and
groups: Learning anticipatory control. J Exp Psych: Learn Mem Cog
2003;{29}:1006--1016.

Krylov NS. Works on the foundations of statistical mechanics.
Princeton Univ Press 1979.

Land, MF and Hayhoe, M. In what ways do eye movements contribute
to everyday activities? Vis Res 2001;41:3559–3565.

Lewin K. Field Theory in Social Science. Univ Chicago Press 1951.

Lewin K. Resolving Social Conflicts and Field Theory in Social
Science. Am Psych Assoc Washington 1997.

Newman-Norlund RD, Noordzij ML, Meulenbroek RGJ and Bekkering H.
Exploring the brain basis of joint action: Co-ordination of
actions goals and intentions. Soc Neurosci\\ 2007;(2)1:48--65.

Pettini M. Geometry and Topology in Hamiltonian Dynamics and
Statistical Mechanics.\\ Springer, New York 2007.

Sch\"{o}ner G. Dynamical Systems Approaches to Cognition. In:
Cambridge Handbook of Computational Cognitive Modelling. Cambridge
Univ Press 2007.

Sebanz N, Bekkering H and Knoblich G. Joint action: bodies and
minds moving together. Tr Cog Sci 2006;(10)2:70--76.

Sutton RS and Barto AG. Reinforcement Learning: An Introduction.
MIT Press, Cambridge 1998.

Todorov E, Jordan MI. Optimal feedback control as a theory of
motor coordination. Nat Neurosci 2002;(5)11:1226-1235.

Tognoli, E, Lagarde, J, DeGuzman, GC and Kelso JAS The phi complex
as a neuromarker of human social coordination. PNAS\\
2007;104(19):8190–8195.

Umezawa H. Advanced field theory: micro macro and thermal
concepts. Am Inst Phys, New York 1993.

Wiener N. Cybernetics. Wiley, New York 1961.

\end{document}

%% file: ivdiag.tex
\newcommand{\mcm}[3]{\newcommand{#1}[#2]{{\ensuremath{#3}}}}

\usepackage{latexsym}
\usepackage{amssymb}

\mcm{\blank}{0}{(\emptybk)} \mcm{\dashbk}{0}{\mbox{---}}
\mcm{\emptybk}{0}{\:\:} \mcm{\hyph}{0}{\mbox{-}}

\mcm{\diagspace}{0}{\mbox{\hspace{2em}}}

\mcm{\cat}{1}{\mc{#1}} \mcm{\fcat}{1}{\mb{#1}}
\mcm{\mc}{1}{\mathcal{#1}} \mcm{\mr}{1}{\mathrm{#1}}
\mcm{\mi}{1}{\mathit{#1}} \mcm{\mb}{1}{\mathbf{#1}}
\mcm{\scat}{1}{\Bbb{#1}} \mcm{\twid}{1}{\widetilde{#1}}

\mcm{\elt}{0}{\in} \mcm{\sub}{0}{\,\subseteq\,}
\mcm{\such}{0}{\:|\:} \mcm{\without}{0}{\setminus}

\mcm{\atsr}{0}{\Box} \mcm{\eqv}{0}{\,\simeq\,}
\mcm{\iso}{0}{\,\cong\,}
\mcm{\of}{0}{\raisebox{0.2mm}{\ensuremath{\scriptstyle\circ}}}

\mcm{\bdry}{0}{\partial}

\mcm{\Bee}{0}{\cat{B}} \mcm{\Beep}{0}{\cat{B'}}
\mcm{\Eee}{0}{\cat{E}} \mcm{\Eeep}{0}{\cat{E'}}

\mcm{\Ess}{0}{\cat{S}} \mcm{\Tee}{0}{\cat{T}}
\mcm{\Teep}{0}{\cat{T'}} \mcm{\Stee}{0}{\scat{T}}
\mcm{\Steep}{0}{\scat{T'}}

\mcm{\blbk}{0}{\blank^{\blob}}
\mcm{\blob}{0}{\scriptscriptstyle{\bullet}}
\mcm{\stbk}{0}{\blank^{*}} \mcm{\ubl}{0}{{}^{\blob}}
\mcm{\ust}{0}{{}^{*}}

\mcm{\Cartpr}{0}{\pr{\Eee}{T}} \mcm{\Cartprp}{0}{\pr{\Eeep}{T'}}
\mcm{\Mnd}{0}{\triple{T}{\eta}{\mu}}
\mcm{\Zeropr}{0}{\pr{\Set}{\id}}

\mcm{\dopset}{0}{\ftrcat{\Delta^{\op}}{\Set}}
\mcm{\tropset}{0}{\ftrcat{\fcat{TR}^{\op}}{\Set}}

\mcm{\cod}{0}{\mr{cod}} \mcm{\dom}{0}{\mr{dom}}
\mcm{\End}{0}{\mr{End}} \mcm{\Hom}{0}{\mr{Hom}}
\mcm{\ob}{0}{\mr{ob}\,} \mcm{\op}{0}{\mr{op}}

\mcm{\comp}{0}{\mi{comp}} \mcm{\id}{0}{\mi{id}}
\mcm{\ids}{0}{\mi{ids}} \mcm{\mult}{0}{\mi{mult}}
\mcm{\unit}{0}{\mi{unit}}

\mcm{\Ab}{0}{\fcat{Ab}} \mcm{\Alg}{0}{\fcat{Alg}}
\mcm{\Bim}{1}{\fcat{Bim}(#1)} \mcm{\Cat}{0}{\fcat{Cat}}
\mcm{\Cay}{0}{\fcat{Cay}} \mcm{\Cpn}{1}{\pr{\Set/S_{#1}}{T_{#1}}}
\mcm{\fc}{0}{\fcat{fc}} \mcm{\fm}{0}{\fcat{fm}}
\mcm{\Graph}{0}{\fcat{Graph}} \mcm{\Gy}{0}{\fcat{Gy}}
\mcm{\Hpn}{1}{\pr{\Eee_{#1}}{P_{#1}}} \mcm{\Mon}{0}{\mb{Mon}}
\mcm{\Multicat}{0}{\fcat{Multicat}} \mcm{\One}{0}{\fcat{1}}
\mcm{\PD}{1}{\fcat{PD}_{#1}} \mcm{\Prof}{0}{\fcat{Prof}}
\mcm{\Set}{0}{\fcat{Set}} \mcm{\Span}{0}{\fcat{Span}}
\mcm{\Ssq}{0}{\fcat{Ssq}} \mcm{\Struc}{0}{\fcat{Struc}}
\mcm{\Sym}{0}{\fcat{Sym}} \mcm{\TR}{1}{\fcat{TR}(#1)}
\mcm{\Tr}{0}{\fcat{Tr}} \mcm{\Twocat}{0}{\fcat{2\hyph\Cat}}

\mcm{\integers}{0}{\mathbb{Z}}

\mcm{\range}{2}{#1,\,\ldots\,,#2}
\mcm{\bftuple}{2}{\tuplebts{\range{#1}{#2}}}
\mcm{\tuple}{3}{\tuplebts{\range{#1,#2}{#3}}}
\mcm{\rttuple}{1}{\tuplebts{\,\ldots\,,#1}}
\mcm{\abftuple}{2}{\atuplebts{\range{#1}{#2}}}
\mcm{\atuple}{3}{\atuplebts{\range{#1,#2}{#3}}}
\mcm{\arttuple}{1}{\atuplebts{\,\ldots\,,#1}}
\mcm{\sqbftuple}{2}{\obt\range{#1}{#2}\cbt}
\mcm{\pr}{2}{\tuplebts{#1,#2}}
\mcm{\triple}{3}{\tuplebts{#1,#2,#3}}

\mcm{\eend}{2}{#1[#2]} \mcm{\ehom}{3}{#1[#2,#3]}
\mcm{\ftrcat}{2}{[#1,#2]} \mcm{\homset}{3}{#1(#2,#3)}
\mcm{\multihom}{3}{#1(#2;#3)}
\mcm{\relhom}{5}{#1_{#2}(\range{#3}{#4};#5)}

\mcm{\go}{0}{\rTo} \mcm{\goby}{1}{\rTo^{#1}}
\mcm{\goesto}{0}{\,\longmapsto\,} \mcm{\goiso}{0}{\goby{\diso}}
\mcm{\monic}{0}{\rMonic} \mcm{\og}{0}{\lTo}
\mcm{\ogby}{1}{\lTo^{#1}}

\mcm{\gph}{2}{\spn{#1}{T #2}{#2}} \mcm{\graph}{4}{\spaan{#1}{T
#2}{#2}{#3}{#4}} \mcm{\oppair}{2}{\stackrel{\rTo^{#1}}{\lTo_{#2}}}
\mcm{\parpair}{2}{\stackrel{\rTo^{#1}}{\rTo_{#2}}}
\mcm{\spn}{3}{#2 \og #1 \go #3} \mcm{\spaan}{5}{#2 \ogby{#4} #1
\goby{#5} #3}

\mcm{\bktdvslob}{3}
    {\left(
    \begin{diagram}[height=1.5em]
    #1      \\
    \dTo>{\,#2} \\
    #3      \\
    \end{diagram}
    \right)}
\mcm{\slob}{3}{(#1 \goby{#2} #3)} \mcm{\vslob}{3}
    {\left.
    \begin{diagram}[height=1.5em]
    #1      \\
    \dTo>{\,#2} \\
    #3      \\
    \end{diagram}
    \right.}

\newenvironment{tree}
    {\begin{diagram}[height=1em,width=.75em,abut,noPS,tight]}
    {\end{diagram}}

\mcm{\enode}{0}{\circ}

\mcm{\nl}{1}{\stackrel{\textstyle #1}{\node}}
\mcm{\node}{0}{\bullet}

\mcm{\utree}{0}{\node}

\mcm{\diso}{0}{\sim}

\mcm{\vdiso}{0}{\wr}

\mcm{\nat}{0}{\mathbb{N}}

\mcm{\Onepr}{0}{\pr{\Graph}{\fc}}
\newlength{\nllwidth}
\newlength{\nllheight}
\newcommand{\stackbelow}[2]{%
\settowidth{\nllwidth}{\ensuremath{#1}\ensuremath{#2}}%
\settoheight{\nllheight}{\ensuremath{#2}}%
\addtolength{\nllheight}{.3ex}%
\mbox{%
\ensuremath{#1}%
\hspace{-.5\nllwidth}%
\raisebox{-1\nllheight}{\ensuremath{#2}}}}
\mcm{\nlal}{2}{\stackbelow{\nl{#1}}{#2}}
\mcm{\nll}{1}{\stackbelow{\node}{#1}} \mcm{\wun}{0}{\fcat{1}}
\mcm{\atuplebts}{1}{\langle #1 \rangle} \mcm{\tuplebts}{1}{(#1)}
\mcm{\bo}{0}{(} \mcm{\bc}{0}{)}
\mcm{\UBilax}{0}{\fcat{UBicat}_\mr{lax}}
\mcm{\UBiwk}{0}{\fcat{UBicat}_\mr{wk}}
\mcm{\UBistr}{0}{\fcat{UBicat}_\mr{str}}
\mcm{\Bilax}{0}{\fcat{Bicat}_\mr{lax}}
\mcm{\Biwk}{0}{\fcat{Bicat}_\mr{wk}}
\mcm{\Bistr}{0}{\fcat{Bicat}_\mr{str}} \mcm{\rotsub}{0}{\cup
\raisebox{0.1em}{$\scriptstyle{|}$}} \mcm{\pd}{0}{\fcat{pd}}
\mcm{\rep}{1}{\widehat{#1}} \mcm{\ovln}{1}{\overline{#1}}
\mcm{\Gph}{0}{\fcat{Gph}} \mcm{\tr}{0}{\fcat{tr}}

\mcm{\ladj}{0}{\,\dashv\,} \mcm{\zeropd}{0}{\node}
    {\end{diagram}}
\mcm{\END}{0}{\fcat{End}} \mcm{\HOM}{0}{\fcat{Hom}}

\newlength{\gwidth} 
\newlength{\gvert}  
\newlength{\gdrop}  
\newlength{\gbaredrop}  
\newlength{\goffset}    
\newlength{\gtemp}  

\newcommand{\present}[1]{%
\makebox[1\gwidth]{%
\rule[-1\gdrop]{0ex}{1\gvert}%
\raisebox{-1\gbaredrop}{#1}}}

\newcommand{\presentl}[1]{%
\makebox[1\gwidth][l]{%
\rule[-1\gdrop]{0ex}{1\gvert}%
\raisebox{-1\gbaredrop}{#1}}}

\newcommand{\presentr}[1]{%
\makebox[1\gwidth][r]{%
\rule[-1\gdrop]{0ex}{1\gvert}%
\raisebox{-1\gbaredrop}{#1}}}

\newcommand{\ginitdims}[2]{
\setlength{\unitlength}{1em}
\setlength{\goffset}{.25\unitlength}
\setlength{\gwidth}{#1\unitlength}
\setlength{\gvert}{#2\unitlength}
\setlength{\gdrop}{.5\gvert}
\addtolength{\gdrop}{-1\goffset}
\setlength{\gbaredrop}{1\gdrop}
\addtolength{\gvert}{.6\unitlength}
\addtolength{\gdrop}{.3\unitlength}}    

\newcommand{\cinitdims}[2]{
\setlength{\unitlength}{1em}
\setlength{\goffset}{.35\unitlength}
\setlength{\gwidth}{#1\unitlength}
\setlength{\gvert}{#2\unitlength}
\setlength{\gdrop}{.5\gvert}
\addtolength{\gdrop}{-1\goffset}
\setlength{\gbaredrop}{1\gdrop}
\addtolength{\gvert}{.6\unitlength}
\addtolength{\gdrop}{.3\unitlength}}    

\newcommand{\gsinitdims}[2]{
\setlength{\unitlength}{0.5em}
\setlength{\goffset}{.25\unitlength}
\setlength{\gwidth}{#1\unitlength}
\setlength{\gvert}{#2\unitlength}
\setlength{\gdrop}{.5\gvert}
\addtolength{\gdrop}{-1\goffset}
\setlength{\gbaredrop}{1\gdrop}
\addtolength{\gvert}{.6\unitlength}
\addtolength{\gdrop}{.3\unitlength}}    

\newcommand{\sidespic}[1]{%
\settowidth{\gtemp}{\ensuremath{#1}}%
\addtolength{\gwidth}{1\gtemp}}

\newcommand{\abovepic}[1]{%
\settoheight{\gtemp}{\ensuremath{#1}}%
\addtolength{\gvert}{1\gtemp}%
\settodepth{\gtemp}{\ensuremath{#1}}%
\addtolength{\gvert}{1\gtemp}}

\newcommand{\belowpic}[1]{%
\settoheight{\gtemp}{\ensuremath{#1}}%
\addtolength{\gvert}{1\gtemp}%
\addtolength{\gdrop}{1\gtemp}%
\settodepth{\gtemp}{\ensuremath{#1}}%
\addtolength{\gvert}{1\gtemp}%
\addtolength{\gdrop}{1\gtemp}}

\newcommand{\cell}[4]{\put(#1,#2){\makebox(0,0)[#3]{\ensuremath{#4}}}}
\mcm{\zmark}{0}{\scriptstyle{\bullet}}

\newcommand{\pregfst}[1]{%
\begin{picture}(0.5,0.2)(-0.5,-0.2)%
\cell{-0.1}{-0.2}{tr}{#1}%
\cell{0}{0}{c}{\zmark}%
\end{picture}}

\mcm{\gfst}{1}{%
\ginitdims{0.5}{0.4}%
\sidespic{#1}%
\belowpic{#1}%
\presentr{\pregfst{#1}}}

\newcommand{\preglst}[1]{%
\begin{picture}(0.5,0.2)(0,-0.2)%
\cell{0.1}{-0.2}{tl}{#1}%
\cell{0.05}{0}{c}{\zmark}%
\end{picture}}

\mcm{\glst}{1}{%
\ginitdims{.5}{.4}%
\sidespic{#1}%
\belowpic{#1}%
\presentl{\preglst{#1}}}

\newcommand{\preglft}[1]{%
\begin{picture}(0,0.2)(0,-0.2)%
\cell{-0.1}{-0.2}{tr}{#1}%
\cell{0.05}{0}{c}{\zmark}%
\end{picture}}

\mcm{\glft}{1}{%
\ginitdims{0}{.4}%
\belowpic{#1}%
\present{\preglft{#1}}}

\newcommand{\pregrgt}[1]{%
\begin{picture}(0,0.2)(0,-0.2)%
\cell{0.1}{-0.2}{tl}{#1}%
\cell{0.05}{0}{c}{\zmark}%
\end{picture}}

\mcm{\grgt}{1}{%
\ginitdims{0}{.4}%
\belowpic{#1}%
\present{\pregrgt{#1}}}

\newcommand{\pregblw}[1]{%
\begin{picture}(0,0.3)(0,-0.3)
\cell{0}{-0.3}{t}{#1}%
\cell{0.05}{0}{c}{\zmark}%
\end{picture}}

\mcm{\gblw}{1}{%
\ginitdims{0}{.6}%
\belowpic{#1}%
\present{\pregblw{#1}}}

\newcommand{\pregfbw}[1]{%
\begin{picture}(0,0.65)(0,-0.65)
\cell{0}{-0.65}{t}{#1}%
\cell{0.05}{0}{c}{\zmark}%
\end{picture}}

\mcm{\gfbw}{1}{%
\ginitdims{0}{1.3}%
\belowpic{#1}%
\present{\pregfbw{#1}}}

\newcommand{\pregzero}[1]{%
\begin{picture}(0.8,0.4)(-0.4,-0.4)
\cell{0}{-0.4}{t}{#1}%
\cell{0}{0}{c}{\zmark}%
\end{picture}}

\mcm{\gzero}{1}{%
\ginitdims{0.8}{.6}%
\belowpic{#1}%
\sidespic{#1}%
\present{\pregzero{#1}}}

\newcommand{\pregone}[1]{%
\begin{picture}(5,0.4)(0,-0.2)%
\cell{2.5}{0.2}{b}{#1}%
\put(0,0){\vector(1,0){5}}%
\end{picture}}

\mcm{\gone}{1}{%
\ginitdims{5}{0.4}%
\abovepic{#1}%
\present{\pregone{#1}}}

\newcommand{\pregtwo}[3]{%
\begin{picture}(5,3.4)(0,-0.2)%
\cell{2.5}{3.2}{b}{#1}%
\cell{2.5}{-.2}{t}{#2}%
\cell{2.7}{1.5}{l}{#3}%
\qbezier(0,1.5)(2.5,4.5)(5,1.5)%
\qbezier(0,1.5)(2.5,-1.5)(5,1.5)%
\put(5,1.5){\vector(1,-1){0}}%
\put(5,1.5){\vector(1,1){0}}%
\put(2.5,2.5){\vector(0,-1){2}}%
\end{picture}}

\mcm{\gtwo}{3}{%
\ginitdims{5}{3.4}%
\abovepic{#1}%
\belowpic{#2}%
\present{\pregtwo{#1}{#2}{#3}}}

\newcommand{\pregthree}[5]{%
\begin{picture}(5,5.4)(0,-1.2)%
\cell{2.5}{4.2}{b}{#1}%
\cell{1.5}{1.7}{b}{#2}%
\cell{2.5}{-1.2}{t}{#3}%
\cell{2.7}{2.75}{l}{#4}%
\cell{2.7}{0.25}{l}{#5}%
\qbezier(0,1.5)(2.5,6.5)(5,1.5)%
\qbezier(0,1.5)(2.5,-3.5)(5,1.5)%
\put(0,1.5){\vector(1,0){5}}%
\put(2.5,3.5){\vector(0,-1){1.5}}%
\put(2.5,1){\vector(0,-1){1.5}}%
\put(5,1.5){\vector(1,-3){0}}%
\put(5,1.5){\vector(1,3){0}}%
\end{picture}}

\mcm{\gthree}{5}{%
\ginitdims{5}{5.4}%
\abovepic{#1}%
\belowpic{#3}%
\present{\pregthree{#1}{#2}{#3}{#4}{#5}}}

\newcommand{\pregfour}[7]{%
\begin{picture}(5,8.4)(0,-2.7)%
\cell{2.5}{5.7}{b}{#1}%
\cell{1.5}{2.8}{b}{#2}%
\cell{1.5}{0.2}{t}{#3}%
\cell{2.5}{-2.7}{t}{#4}%
\cell{2.7}{4.25}{l}{#5}%
\cell{2.7}{1.5}{l}{#6}%
\cell{2.7}{-1.25}{l}{#7}%
\qbezier(0,1.5)(2.5,9.5)(5,1.5)%
\qbezier(0,1.5)(2.5,4)(5,1.5)%
\qbezier(0,1.5)(2.5,-1)(5,1.5)%
\qbezier(0,1.5)(2.5,-6.5)(5,1.5)%
\put(2.5,5.25){\vector(0,-1){2}}%
\put(2.5,2.5){\vector(0,-1){2}}%
\put(2.5,-0.25){\vector(0,-1){2}}%
\put(5,1.5){\vector(1,-4){0}}%
\put(5,1.5){\vector(4,-3){0}}%
\put(5,1.5){\vector(4,3){0}}%
\put(5,1.5){\vector(1,4){0}}%
\end{picture}}

\mcm{\gfour}{7}{%
\ginitdims{5}{8.4}%
\abovepic{#1}%
\belowpic{#4}%
\present{\pregfour{#1}{#2}{#3}{#4}{#5}{#6}{#7}}}

\newcommand{\pregthreecell}[5]{%
\begin{picture}(8,5)(-4,-2.5)%
\cell{0}{2.5}{b}{#1}%
\cell{0}{-2.5}{t}{#2}%
\cell{-1.7}{0}{r}{#3}%
\cell{1.7}{0}{l}{#4}%
\cell{0}{0.2}{b}{#5}%
\qbezier(-4,0)(0,4.2)(4,0)%
\qbezier(-4,0)(0,-4.2)(4,0)%
\qbezier(-0.5,1.8)(-2.5,0)(-0.5,-1.8)%
\qbezier(0.5,1.8)(2.5,0)(0.5,-1.8)%
\put(-1,0){\vector(1,0){2}}%
\put(4,0){\vector(1,-1){0}}%
\put(4,0){\vector(1,1){0}}%
\put(-0.5,-1.8){\vector(1,-1){0}}%
\put(0.5,-1.8){\vector(-1,-1){0}}%
\end{picture}}

\mcm{\gthreecell}{5}{%
\ginitdims{8}{5}%
\abovepic{#1}%
\belowpic{#2}%
\present{\pregthreecell{#1}{#2}{#3}{#4}{#5}}}

\newcommand{\pregthreecellu}{%
\begin{picture}(5,3.4)(-0.5,-0.2)%
\qbezier(-.5,1.5)(2,4.5)(4.5,1.5)%
\qbezier(-.5,1.5)(2,-1.5)(4.5,1.5)%
\qbezier(1.5,2.7)(0.5,1.5)(1.5,0.3)%
\qbezier(2.5,2.7)(3.5,1.5)(2.5,0.3)%
\put(1.3,1.5){\vector(1,0){1.4}}%
\put(4.5,1.5){\vector(1,-1){0}}%
\put(4.5,1.5){\vector(1,1){0}}%
\put(1.5,0.3){\vector(2,-3){0}}%
\put(2.5,0.3){\vector(-2,-3){0}}%
\end{picture}}

\mcm{\gthreecellu}{0}{%
\ginitdims{5}{3.4}%
\present{\pregthreecellu}}

\newcommand{\pregtwocentre}[3]{%
\begin{picture}(5,3.4)(0,-0.2)%
\cell{2.5}{3.2}{b}{#1}%
\cell{2.5}{-.2}{t}{#2}%
\cell{2.5}{1.5}{c}{#3}%
\qbezier(0,1.5)(2.5,4.5)(5,1.5)%
\qbezier(0,1.5)(2.5,-1.5)(5,1.5)%
\put(5,1.5){\vector(1,-1){0}}%
\put(5,1.5){\vector(1,1){0}}%
\put(2.5,2.5){\vector(0,-1){2}}%
\end{picture}}

\mcm{\gtwocentre}{3}{%
\ginitdims{5}{3.4}%
\abovepic{#1}%
\belowpic{#2}%
\present{\pregtwocentre{#1}{#2}{#3}}}

\newcommand{\pregspecialone}[9]{%
\begin{picture}(8,8)(-4,-4)%
\cell{0}{3.9}{b}{#1}%
\cell{-2}{-0.2}{t}{#2}%
\cell{0}{-3.9}{t}{#3}%
\cell{-1.5}{1.1}{r}{#4}%
\cell{0.2}{1.5}{l}{#5}%
\cell{1.5}{1.1}{l}{#6}%
\cell{0.2}{-2}{l}{#7}%
\cell{-0.9}{2.3}{b}{#8}%
\cell{0.9}{2.3}{b}{#9}%
\qbezier(-4,0)(0,8)(4,0)%
\qbezier(-4,0)(0,-8)(4,0)%
\qbezier(-0.5,3.4)(-3.5,2)(-0.5,0.6)%
\qbezier(0.5,3.4)(3.5,2)(0.5,0.6)%
\put(-4,0){\vector(1,0){8}}%
\put(0,3.4){\vector(0,-1){2.8}}%
\put(0,-0.8){\vector(0,-1){2.4}}%
\put(-1.5,2.2){\vector(1,0){1.2}}%
\put(0.3,2.2){\vector(1,0){1.2}}%
\put(4,0){\vector(1,-2){0}}%
\put(4,0){\vector(1,2){0}}%
\put(-0.5,0.6){\vector(2,-1){0}}%
\put(0.5,0.6){\vector(-2,-1){0}}%
\end{picture}}

\mcm{\gspecialone}{9}{%
\ginitdims{8}{8}%
\abovepic{#1}%
\belowpic{#3}%
\present{\pregspecialone{#1}{#2}{#3}{#4}{#5}{#6}{#7}{#8}{#9}}}

\newcommand{\pregspecialtwo}{%
\begin{picture}(5,3.4)(0,-0.2)%
\qbezier(0,1.5)(2.5,4.5)(5,1.5)%
\qbezier(0,1.5)(2.5,-1.5)(5,1.5)%
\qbezier(1.7,2.5)(0,1.5)(1.7,0.5)%
\qbezier(3.3,2.5)(5,1.5)(3.3,0.5)%
\put(5,1.5){\vector(1,-1){0}}%
\put(5,1.5){\vector(1,1){0}}%
\put(1.7,0.5){\vector(3,-2){0}}%
\put(3.3,0.5){\vector(-3,-2){0}}%
\put(2.5,2.5){\vector(0,-1){2}}%
\put(1.2,1.5){\vector(1,0){1}}%
\put(2.8,1.5){\vector(1,0){1}}%
\end{picture}}

\mcm{\gspecialtwo}{0}{%
\ginitdims{5}{3.4}%
\present{\pregspecialtwo}}

\newcommand{\pregspecialthree}{%
\begin{picture}(5,5.4)(0,-1.2)%
\qbezier(0,1.5)(2.5,6.5)(5,1.5)%
\qbezier(0,1.5)(2.5,-3.5)(5,1.5)%
\qbezier(2,3.5)(1,2.75)(2,2)%
\qbezier(3,3.5)(4,2.75)(3,2)%
\qbezier(2,1)(1,0.25)(2,-0.5)%
\qbezier(3,1)(4,0.25)(3,-0.5)%
\put(0,1.5){\vector(1,0){5}}%
\put(1.5,2.75){\vector(1,0){2}}%
\put(1.5,0.25){\vector(1,0){2}}%
\put(5,1.5){\vector(1,-3){0}}%
\put(5,1.5){\vector(1,3){0}}%
\put(2,2){\vector(1,-1){0}}%
\put(3,2){\vector(-1,-1){0}}%
\put(2,-0.5){\vector(1,-1){0}}%
\put(3,-0.5){\vector(-1,-1){0}}%
\end{picture}}

\mcm{\gspecialthree}{0}{%
\ginitdims{5}{5.4}%
\present{\pregspecialthree}}

\newcommand{\pregonew}[1]{%
\begin{picture}(8,0.4)(0,-0.2)%
\cell{4}{0.2}{b}{#1}%
\put(0,0){\vector(1,0){8}}%
\end{picture}}

\mcm{\gonew}{1}{%
\ginitdims{8}{0.4}%
\abovepic{#1}%
\present{\pregonew{#1}}}

\mcm{\gzersu}{0}{%
\gsinitdims{0}{.6}%
\present{\pregblw{}}}

\mcm{\gonesu}{0}{%
\gsinitdims{5}{0.4}%
\present{\pregone{}}}

\mcm{\gtwosu}{0}{%
\gsinitdims{5}{3.4}%
\present{\pregtwo{}{}{}}}

\mcm{\gthreesu}{0}{%
\gsinitdims{5}{5.4}%
\present{\pregthree{}{}{}{}{}}}

\mcm{\gfoursu}{0}{%
\gsinitdims{5}{8.4}%
\present{\pregfour{}{}{}{}{}{}{}}}

\newcommand{\precone}[1]{%
\begin{picture}(4.2,0.4)(-0.3,-0.2)%
\cell{1.8}{0.2}{b}{#1}%
\put(0,0){\vector(1,0){3.6}}%
\end{picture}}

\mcm{\cone}{1}{%
\cinitdims{4.2}{0.4}%
\abovepic{#1}%
\present{\precone{#1}}}

\mcm{\gfstsu}{0}{%
\gsinitdims{0.5}{0.4}%
\presentr{\pregfst{}}}

\mcm{\glstsu}{0}{%
\gsinitdims{0.5}{0.4}%
\presentl{\preglst{}}}

\newcommand{\prectwodbl}[3]%
{\begin{picture}(4.2,3.4)(-0.1,-0.2)%
\cell{2}{3.2}{b}{#1}%
\cell{2}{-0.2}{t}{#2}%
\cell{2.3}{1.5}{l}{#3}%
\qbezier(0,2)(2,4)(4,2)%
\qbezier(0,1)(2,-1)(4,1)%
\put(4,2){\vector(1,-1){0}}%
\put(4,1){\vector(1,1){0}}%
\put(1.9,2.5){\line(0,-1){1.8}}%
\put(2.1,2.5){\line(0,-1){1.8}}%
\cell{2.01}{0.4}{b}{\vee}%
\end{picture}}

\mcm{\ctwodbl}{3}{%
\cinitdims{4.2}{3.4}%
\abovepic{#1}%
\belowpic{#2}%
\present{\prectwodbl{#1}{#2}{#3}}}

\newcommand{\precthreedbl}[5]{%
\begin{picture}(4.2,5.4)(-0.1,-0.2)%
\cell{2}{5.2}{b}{#1}%
\cell{1}{2.7}{b}{#2}%
\cell{2}{-.2}{t}{#3}%
\cell{2.3}{3.75}{l}{#4}%
\cell{2.3}{1.25}{l}{#5}%
\qbezier(0,3)(2,7)(4,3)%
\qbezier(0,2)(2,-2)(4,2)%
\put(0,2.5){\vector(1,0){4}}%
\put(1.9,4.5){\line(0,-1){1.3}}%
\put(2.1,4.5){\line(0,-1){1.3}}%
\cell{2.01}{2.9}{b}{\vee}%
\put(1.9,2){\line(0,-1){1.3}}%
\put(2.1,2){\line(0,-1){1.3}}%
\cell{2.01}{0.4}{b}{\vee}%
\put(4,3){\vector(1,-3){0}}%
\put(4,2){\vector(1,3){0}}%
\end{picture}}

\mcm{\cthreedbl}{5}{%
\cinitdims{4.2}{5.4}%
\abovepic{#1}%
\belowpic{#3}%
\present{\precthreedbl{#1}{#2}{#3}{#4}{#5}}}

\newcommand{\precthreecelltrp}[5]{%
\begin{picture}(8.2,5)(-4.1,-2.5)%
\cell{0}{2.5}{b}{#1}%
\cell{0}{-2.5}{t}{#2}%
\cell{-1.8}{0}{r}{#3}%
\cell{1.8}{0}{l}{#4}%
\cell{0}{0.3}{b}{#5}%
\qbezier(-4,0.5)(0,4)(4,0.5)%
\qbezier(-4,-0.5)(0,-4)(4,-0.5)%
\qbezier(-0.6,2)(-2.6,0)(-0.6,-2)%
\qbezier(-0.4,2)(-2.4,0)(-0.5,-1.9)%
\cell{-0.6}{-2}{b}{\lrcorner}%
\qbezier(0.4,2)(2.4,0)(0.5,-1.9)%
\qbezier(0.6,2)(2.6,0)(0.6,-2)%
\cell{0.65}{-2}{b}{\llcorner}%
\put(-1,0.15){\line(1,0){1.7}}%
\put(-1,0){\line(1,0){2}}%
\put(-1,-0.15){\line(1,0){1.7}}%
\cell{1.15}{0}{r}{>}%
\put(4,0.5){\vector(1,-1){0}}%
\put(4,-0.5){\vector(1,1){0}}%
\end{picture}}

\mcm{\cthreecelltrp}{5}{%
\cinitdims{8.2}{5}%
\abovepic{#1}%
\belowpic{#2}%
\present{\precthreecelltrp{#1}{#2}{#3}{#4}{#5}}}

\newcommand{\prectwo}[3]%
{\begin{picture}(4.2,3.4)(-0.1,-0.2)%
\cell{2}{3.2}{b}{#1}%
\cell{2}{-0.2}{t}{#2}%
\cell{2.2}{1.5}{l}{#3}%
\qbezier(0,2)(2,4)(4,2)%
\qbezier(0,1)(2,-1)(4,1)%
\put(4,2){\vector(1,-1){0}}%
\put(4,1){\vector(1,1){0}}%
\put(2,2.5){\vector(0,-1){2}}%
\end{picture}}

\mcm{\ctwo}{3}{%
\cinitdims{4.2}{3.4}%
\abovepic{#1}%
\belowpic{#2}%
\present{\prectwo{#1}{#2}{#3}}}

\newcommand{\precthree}[5]{%
\begin{picture}(4.2,5.4)(-0.1,-0.2)%
\cell{2}{5.2}{b}{#1}%
\cell{1}{2.7}{b}{#2}%
\cell{2}{-.2}{t}{#3}%
\cell{2.2}{3.75}{l}{#4}%
\cell{2.2}{1.25}{l}{#5}%
\qbezier(0,3)(2,7)(4,3)%
\qbezier(0,2)(2,-2)(4,2)%
\put(0,2.5){\vector(1,0){4}}%
\put(2,4.5){\vector(0,-1){1.5}}%
\put(2,2){\vector(0,-1){1.5}}%
\put(4,3){\vector(1,-3){0}}%
\put(4,2){\vector(1,3){0}}%
\end{picture}}

\mcm{\cthree}{5}{%
\cinitdims{4.2}{5.4}%
\abovepic{#1}%
\belowpic{#3}%
\present{\precthree{#1}{#2}{#3}{#4}{#5}}}

\newcommand{\prectwoop}[3]%
{\begin{picture}(4.2,3.4)(-0.1,-0.2)%
\cell{2}{3.2}{b}{#1}%
\cell{2}{-0.2}{t}{#2}%
\cell{2.2}{1.5}{l}{#3}%
\qbezier(0,2)(2,4)(4,2)%
\qbezier(0,1)(2,-1)(4,1)%
\put(0,2){\vector(-1,-1){0}}%
\put(0,1){\vector(-1,1){0}}%
\put(2,2.5){\vector(0,-1){2}}%
\end{picture}}

\mcm{\ctwoop}{3}{%
\cinitdims{4.2}{3.4}%
\abovepic{#1}%
\belowpic{#2}%
\present{\prectwoop{#1}{#2}{#3}}}

\newcommand{\prectwopar}[4]{%
\begin{picture}(4.2,3.4)(-0.1,-0.2)%
\cell{2}{3.2}{b}{#1}%
\cell{2}{-0.2}{t}{#2}%
\cell{1.6}{1.5}{r}{#3}%
\cell{2.4}{1.5}{l}{#4}%
\qbezier(0,2)(2,4)(4,2)%
\qbezier(0,1)(2,-1)(4,1)%
\put(4,2){\vector(1,-1){0}}%
\put(4,1){\vector(1,1){0}}%
\put(1.8,2.5){\vector(0,-1){2}}%
\put(2.2,2.5){\vector(0,-1){2}}%
\end{picture}}

\mcm{\ctwopar}{4}{%
\cinitdims{4.2}{3.4}%
\abovepic{#1}%
\belowpic{#2}%
\present{\prectwopar{#1}{#2}{#3}{#4}}}

\newcommand{\precthreein}[5]{%
\begin{picture}(4.2,5.4)(-0.1,-0.2)%
\cell{2}{5.2}{b}{#1}%
\cell{1}{2.7}{b}{#2}%
\cell{2}{-.2}{t}{#3}%
\cell{2.2}{3.75}{l}{#4}%
\cell{2.2}{1.25}{l}{#5}%
\qbezier(0,3)(2,7)(4,3)%
\qbezier(0,2)(2,-2)(4,2)%
\put(0,2.5){\vector(1,0){4}}%
\put(2,4.5){\vector(0,-1){1.5}}%
\put(2,0.5){\vector(0,1){1.5}}%
\put(4,3){\vector(1,-3){0}}%
\put(4,2){\vector(1,3){0}}%
\end{picture}}

\mcm{\cthreein}{5}{%
\cinitdims{4.2}{5.4}%
\abovepic{#1}%
\belowpic{#3}%
\present{\precthreein{#1}{#2}{#3}{#4}{#5}}}

\newcommand{\precthreecell}[5]{%
\begin{picture}(8.2,5)(-4.1,-2.5)%
\cell{0}{2.5}{b}{#1}%
\cell{0}{-2.5}{t}{#2}%
\cell{-1.7}{0}{r}{#3}%
\cell{1.7}{0}{l}{#4}%
\cell{0}{0.2}{b}{#5}%
\qbezier(-4,0.5)(0,4)(4,0.5)%
\qbezier(-4,-0.5)(0,-4)(4,-0.5)%
\qbezier(-0.5,2)(-2.5,0)(-0.5,-2)%
\qbezier(0.5,2)(2.5,0)(0.5,-2)%
\put(-1,0){\vector(1,0){2}}%
\put(4,0.5){\vector(1,-1){0}}%
\put(4,-0.5){\vector(1,1){0}}%
\put(-0.5,-2){\vector(1,-1){0}}%
\put(0.5,-2){\vector(-1,-1){0}}%
\end{picture}}

\mcm{\cthreecell}{5}{%
\cinitdims{8.2}{5}%
\abovepic{#1}%
\belowpic{#2}%
\present{\precthreecell{#1}{#2}{#3}{#4}{#5}}}

\newcommand{\precthreecellpar}[6]{%
\begin{picture}(8.2,5)(-4.1,-2.5)%
\cell{0}{2.5}{b}{#1}%
\cell{0}{-2.5}{t}{#2}%
\cell{-1.7}{0}{r}{#3}%
\cell{1.7}{0}{l}{#4}%
\cell{0}{0.4}{b}{#5}%
\cell{0}{-0.4}{t}{#6}%
\qbezier(-4,0.5)(0,4)(4,0.5)%
\qbezier(-4,-0.5)(0,-4)(4,-0.5)%
\qbezier(-0.5,2)(-2.5,0)(-0.5,-2)%
\qbezier(0.5,2)(2.5,0)(0.5,-2)%
\put(-1,0.2){\vector(1,0){2}}%
\put(-1,-0.2){\vector(1,0){2}}%
\put(4,0.5){\vector(1,-1){0}}%
\put(4,-0.5){\vector(1,1){0}}%
\put(-0.5,-2){\vector(1,-1){0}}%
\put(0.5,-2){\vector(-1,-1){0}}%
\end{picture}}

\mcm{\cthreecellpar}{6}{%
\cinitdims{8.2}{5}%
\abovepic{#1}%
\belowpic{#2}%
\present{\precthreecellpar{#1}{#2}{#3}{#4}{#5}{#6}}}

\newcommand{\prectwov}[5]{%
\begin{picture}(3.4,4.2)(0.8,0.9)%
\cell{2.5}{5.1}{b}{#1}%
\cell{2.5}{0.9}{t}{#2}%
\cell{0.8}{3}{r}{#3}%
\cell{4.2}{3}{l}{#4}%
\cell{2.5}{3.2}{b}{#5}%
\qbezier(2,5)(0,3)(2,1)%
\qbezier(3,5)(5,3)(3,1)%
\put(2,1){\vector(1,-1){0}}%
\put(3,1){\vector(-1,-1){0}}%
\put(1.5,3){\vector(1,0){2}}%
\end{picture}}

\mcm{\ctwov}{5}{%
\cinitdims{3.4}{4.2}%
\abovepic{#1}%
\belowpic{#2}%
\sidespic{#3}%
\sidespic{#4}%
\present{\prectwov{#1}{#2}{#3}{#4}{#5}}}

\newcommand{\precthreecellv}[7]{%
\begin{picture}(5,8.2)(0.5,-1.6)%
\cell{3}{6.6}{b}{#1}%
\cell{3}{-1.6}{t}{#2}%
\cell{0.5}{2.5}{r}{#3}%
\cell{5.5}{2.5}{l}{#4}%
\cell{3}{4.2}{b}{#5}%
\cell{3}{0.8}{t}{#6}%
\cell{3.2}{2.5}{l}{#7}%
\qbezier(3.5,6.5)(7,2.5)(3.5,-1.5)%
\qbezier(2.5,6.5)(-1,2.5)(2.5,-1.5)%
\put(2.5,-1.5){\vector(1,-1){0}}%
\put(3.5,-1.5){\vector(-1,-1){0}}%
\qbezier(1,3)(3,5)(5,3)%
\qbezier(1,2)(3,0)(5,2)%
\put(5,3){\vector(1,-1){0}}%
\put(5,2){\vector(1,1){0}}%
\put(3,3.5){\vector(0,-1){2}}%
\end{picture}}

\mcm{\cthreecellv}{7}{%
\cinitdims{5}{8.2}%
\abovepic{#1}%
\belowpic{#2}%
\sidespic{#3}%
\sidespic{#4}%
\present{\precthreecellv{#1}{#2}{#3}{#4}{#5}{#6}{#7}}}

\newcommand{\pretopez}[2]{%
\begin{picture}(2.6,2.3)(-1.3,-2.2)%
\cell{0}{-2.2}{t}{#1}%
\cell{0}{-1.2}{c}{#2}%
\qbezier(0,0)(-2,-2)(0,-2)%
\qbezier(0,0)(2,-2)(0,-2)%
\put(0,0){\vector(-1,1){0}}%
\end{picture}}

\mcm{\topez}{2}{%
\ginitdims{2.6}{2.3}%
\belowpic{#1}%
\present{\pretopez{#1}{#2}}}

\newcommand{\pretopea}[3]{%
\begin{picture}(4,1.9)(-2,-0,2)%
\cell{0}{1.7}{b}{#1}%
\cell{0}{-0.2}{t}{#2}%
\cell{0}{0.7}{c}{#3}%
\qbezier(-2,0)(0,3)(2,0)%
\put(-2,0){\vector(1,0){4}}%
\put(2,0){\vector(2,-3){0}}%
\end{picture}}

\mcm{\topea}{3}{%
\ginitdims{4}{1.9}%
\abovepic{#1}%
\belowpic{#2}%
\present{\pretopea{#1}{#2}{#3}}}

\newcommand{\pretopeb}[4]{%
\begin{picture}(4,2.2)(-2,-0.2)%
\cell{-1.1}{1}{br}{#1}%
\cell{1.1}{1}{bl}{#2}%
\cell{0}{-0.2}{t}{#3}%
\cell{0}{0.8}{c}{#4}%
\put(-2,0){\vector(1,1){2}}%
\put(0,2){\vector(1,-1){2}}%
\put(-2,0){\vector(1,0){4}}%
\end{picture}}

\mcm{\topeb}{4}{%
\ginitdims{4}{2.2}%
\belowpic{#3}%
\present{\pretopeb{#1}{#2}{#3}{#4}}}

\newcommand{\pretopec}[5]{%
\begin{picture}(4,2.2)(-2,-0.2)%
\cell{-1.8}{1}{br}{#1}%
\cell{0}{2.2}{b}{#2}%
\cell{1.8}{1}{bl}{#3}%
\cell{0}{-0.2}{t}{#4}%
\cell{0}{0.8}{c}{#5}%
\put(-2,0){\vector(1,2){1}}%
\put(-1,2){\vector(1,0){2}}%
\put(1,2){\vector(1,-2){1}}%
\put(-2,0){\vector(1,0){4}}%
\end{picture}}

\mcm{\topec}{5}{%
\ginitdims{4}{2.2}%
\sidespic{#1}%
\abovepic{#2}%
\sidespic{#3}%
\belowpic{#4}%
\present{\pretopec{#1}{#2}{#3}{#4}{#5}}}

\newcommand{\pretoped}[6]{%
\begin{picture}(4,2.5)(-2,-0.2)%
\cell{-2}{0.6}{br}{#1}%
\cell{-0.7}{2.2}{br}{#2}%
\cell{0.7}{2.2}{bl}{#3}%
\cell{2}{0.6}{bl}{#4}%
\cell{0}{-0.2}{t}{#5}%
\cell{0}{0.8}{c}{#6}%
\put(-2,0){\vector(1,3){0.5}}%
\put(-1.5,1.5){\vector(3,2){1.5}}%
\put(0,2.5){\vector(3,-2){1.5}}%
\put(1.5,1.5){\vector(1,-3){0.5}}%
\put(-2,0){\vector(1,0){4}}%
\end{picture}}

\mcm{\toped}{6}{%
\ginitdims{4}{2.5}%
\sidespic{#1}%
\abovepic{#2}%
\abovepic{#3}%
\sidespic{#4}%
\belowpic{#5}%
\present{\pretoped{#1}{#2}{#3}{#4}{#5}{#6}}}

\newcommand{\pretopeq}[5]{%
\begin{picture}(4,2.5)(-2,-0.2)%
\cell{-2}{0.6}{br}{#1}%
\cell{-1}{2.2}{br}{#2}%
\cell{2}{0.6}{bl}{#3}%
\cell{0}{-0.2}{t}{#4}%
\cell{0}{0.8}{c}{#5}%
\put(-2,0){\vector(1,3){0.5}}%
\put(-1.5,1.5){\vector(1,1){1}}%
\cell{0.9}{2.3}{c}{\ddots}
\put(1.5,1.5){\vector(1,-3){0.5}}%
\put(-2,0){\vector(1,0){4}}%
\end{picture}}

\mcm{\topeq}{5}{%
\ginitdims{4}{2.5}%
\sidespic{#1}%
\abovepic{#2}%
\sidespic{#3}%
\belowpic{#4}%
\present{\pretopeq{#1}{#2}{#3}{#4}{#5}}}

\newcommand{\pretopebase}[1]{%
\begin{picture}(4,0.4)(0,-0.2)%
\cell{2}{0.2}{b}{#1}%
\put(0,0){\vector(1,0){4}}%
\end{picture}}

\mcm{\topebase}{1}{%
\ginitdims{4}{0.4}%
\abovepic{#1}%
\present{\pretopebase{#1}}}

\newcommand{\pretopezs}[2]{%
\begin{picture}(2.6,2.3)(-1.3,-2.2)%
\cell{0}{-2.2}{t}{#1}%
\cell{0}{-1.2}{c}{#2}%
\qbezier(0,0)(-2,-2)(0,-2)%
\qbezier(0,0)(2,-2)(0,-2)%
\end{picture}}

\mcm{\topezs}{2}{%
\ginitdims{2.6}{2.3}%
\belowpic{#1}%
\present{\pretopezs{#1}{#2}}}

\newcommand{\pretopeas}[3]{%
\begin{picture}(4,1.9)(-2,-0,2)%
\cell{0}{1.7}{b}{#1}%
\cell{0}{-0.2}{t}{#2}%
\cell{0}{0.7}{c}{#3}%
\qbezier(-2,0)(0,3)(2,0)%
\put(-2,0){\line(1,0){4}}%
\end{picture}}

\mcm{\topeas}{3}{%
\ginitdims{4}{1.9}%
\abovepic{#1}%
\belowpic{#2}%
\present{\pretopeas{#1}{#2}{#3}}}

\newcommand{\pretopebs}[4]{%
\begin{picture}(4,2.2)(-2,-0.2)%
\cell{-1.1}{1}{br}{#1}%
\cell{1.1}{1}{bl}{#2}%
\cell{0}{-0.2}{t}{#3}%
\cell{0}{0.8}{c}{#4}%
\put(-2,0){\line(1,1){2}}%
\put(0,2){\line(1,-1){2}}%
\put(-2,0){\line(1,0){4}}%
\end{picture}}

\mcm{\topebs}{4}{%
\ginitdims{4}{2.2}%
\belowpic{#3}%
\present{\pretopebs{#1}{#2}{#3}{#4}}}

\newcommand{\pretopecs}[5]{%
\begin{picture}(4,2.2)(-2,-0.2)%
\cell{-1.8}{1}{br}{#1}%
\cell{0}{2.2}{b}{#2}%
\cell{1.8}{1}{bl}{#3}%
\cell{0}{-0.2}{t}{#4}%
\cell{0}{0.8}{c}{#5}%
\put(-2,0){\line(1,2){1}}%
\put(-1,2){\line(1,0){2}}%
\put(1,2){\line(1,-2){1}}%
\put(-2,0){\line(1,0){4}}%
\end{picture}}

\mcm{\topecs}{5}{%
\ginitdims{4}{2.2}%
\sidespic{#1}%
\abovepic{#2}%
\sidespic{#3}%
\belowpic{#4}%
\present{\pretopecs{#1}{#2}{#3}{#4}{#5}}}

\newcommand{\pretopeds}[6]{%
\begin{picture}(4,2.5)(-2,-0.2)%
\cell{-2}{0.6}{br}{#1}%
\cell{-0.7}{2.2}{br}{#2}%
\cell{0.7}{2.2}{bl}{#3}%
\cell{2}{0.6}{bl}{#4}%
\cell{0}{-0.2}{t}{#5}%
\cell{0}{0.8}{c}{#6}%
\put(-2,0){\line(1,3){0.5}}%
\put(-1.5,1.5){\line(3,2){1.5}}%
\put(0,2.5){\line(3,-2){1.5}}%
\put(1.5,1.5){\line(1,-3){0.5}}%
\put(-2,0){\line(1,0){4}}%
\end{picture}}

\mcm{\topeds}{6}{%
\ginitdims{4}{2.5}%
\sidespic{#1}%
\abovepic{#2}%
\abovepic{#3}%
\sidespic{#4}%
\belowpic{#5}%
\present{\pretopeds{#1}{#2}{#3}{#4}{#5}{#6}}}

\newcommand{\pretopeqs}[5]{%
\begin{picture}(4,2.5)(-2,-0.2)%
\cell{-2}{0.6}{br}{#1}%
\cell{-1}{2.2}{br}{#2}%
\cell{2}{0.6}{bl}{#3}%
\cell{0}{-0.2}{t}{#4}%
\cell{0}{0.8}{c}{#5}%
\put(-2,0){\line(1,3){0.5}}%
\put(-1.5,1.5){\line(1,1){1}}%
\cell{0.9}{2.3}{c}{\ddots}
\put(1.5,1.5){\line(1,-3){0.5}}%
\put(-2,0){\line(1,0){4}}%
\end{picture}}

\mcm{\topeqs}{5}{%
\ginitdims{4}{2.5}%
\sidespic{#1}%
\abovepic{#2}%
\sidespic{#3}%
\belowpic{#4}%
\present{\pretopeqs{#1}{#2}{#3}{#4}{#5}}}

\newcommand{\pretopebases}[1]{%
\begin{picture}(4,0.4)(0,-0.2)%
\cell{2}{0.2}{b}{#1}%
\put(0,0){\line(1,0){4}}%
\end{picture}}

\mcm{\topebases}{1}{%
\ginitdims{4}{0.4}%
\abovepic{#1}%
\present{\pretopebases{#1}}}

\newcommand{\pregdots}[6]{%
\begin{picture}(5,8.4)(0,-2.7)%
\cell{2.5}{5.7}{b}{#1}%
\cell{1.5}{2.8}{b}{#2}%
\cell{1.5}{0.2}{t}{#3}%
\cell{2.5}{-2.7}{t}{#4}%
\cell{2.7}{4.25}{l}{#5}%
\cell{2.7}{-1.25}{l}{#6}%
\qbezier(0,1.5)(2.5,9.5)(5,1.5)%
\qbezier(0,1.5)(2.5,4)(5,1.5)%
\qbezier(0,1.5)(2.5,-1)(5,1.5)%
\qbezier(0,1.5)(2.5,-6.5)(5,1.5)%
\put(2.5,5.25){\vector(0,-1){2}}%
\put(2.5,-0.25){\vector(0,-1){2}}%
\cell{2.5}{1.7}{c}{\vdots}%
\put(5,1.5){\vector(1,-4){0}}%
\put(5,1.5){\vector(4,-3){0}}%
\put(5,1.5){\vector(4,3){0}}%
\put(5,1.5){\vector(1,4){0}}%
\end{picture}}

\mcm{\gdots}{6}{%
\ginitdims{5}{8.4}%
\abovepic{#1}%
\belowpic{#4}%
\present{\pregdots{#1}{#2}{#3}{#4}{#5}{#6}}}

\newlength{\volt}
\makeatletter

\def\diagram{\m@th\leftwidth=\z@ \rightwidth=\z@ \topheight=\z@
\botheight=\z@ \setbox\@picbox\hbox\bgroup}

\def\enddiagram{\egroup\wd\@picbox\rightwidth\unitlength
\ht\@picbox\topheight\unitlength \dp\@picbox\botheight\unitlength
\hskip\leftwidth\unitlength\box\@picbox}

\def\bfig{\begin{diagram}}
\def\efig{\end{diagram}}
\newcount\wideness \newcount\leftwidth \newcount\rightwidth
\newcount\highness \newcount\topheight \newcount\botheight

\def\ratchet#1#2{\ifnum#1<#2 \global #1=#2 \fi}

\def\putbox(#1,#2)#3{%
\horsize{\wideness}{#3} \divide\wideness by 2 {\advance\wideness
by #1 \ratchet{\rightwidth}{\wideness}} {\advance\wideness by -#1
\ratchet{\leftwidth}{\wideness}} \vertsize{\highness}{#3}
\divide\highness by 2 {\advance\highness by #2
\ratchet{\topheight}{\highness}} {\advance\highness by -#2
\ratchet{\botheight}{\highness}} \put(#1,#2){\makebox(0,0){$#3$}}}

\def\putlbox(#1,#2)#3{%
\horsize{\wideness}{#3} {\advance\wideness by #1
\ratchet{\rightwidth}{\wideness}} {\ratchet{\leftwidth}{-#1}}
\vertsize{\highness}{#3} \divide\highness by 2 {\advance\highness
by #2 \ratchet{\topheight}{\highness}} {\advance\highness by -#2
\ratchet{\botheight}{\highness}}
\put(#1,#2){\makebox(0,0)[l]{$#3$}}}

\def\putrbox(#1,#2)#3{%
\horsize{\wideness}{#3} {\ratchet{\rightwidth}{#1}}
{\advance\wideness by -#1 \ratchet{\leftwidth}{\wideness}}
\vertsize{\highness}{#3} \divide\highness by 2 {\advance\highness
by #2 \ratchet{\topheight}{\highness}} {\advance\highness by -#2
\ratchet{\botheight}{\highness}}
\put(#1,#2){\makebox(0,0)[r]{$#3$}}}

\def\adjust[#1]{} 

\newcount \coefa
\newcount \coefb
\newcount \coefc
\newcount\tempcounta
\newcount\tempcountb
\newcount\tempcountc
\newcount\tempcountd
\newcount\xext
\newcount\yext
\newcount\xoff
\newcount\yoff
\newcount\gap%
\newcount\arrowtypea
\newcount\arrowtypeb
\newcount\arrowtypec
\newcount\arrowtyped
\newcount\arrowtypee
\newcount\height
\newcount\width
\newcount\xpos
\newcount\ypos
\newcount\run
\newcount\rise
\newcount\arrowlength
\newcount\halflength
\newcount\arrowtype
\newdimen\tempdimen
\newdimen\xlen
\newdimen\ylen
\newsavebox{\tempboxa}%
\newsavebox{\tempboxb}%
\newsavebox{\tempboxc}%

\newdimen\w@dth

\def\setw@dth#1#2{\setbox\z@\hbox{\m@th$#1$}\w@dth=\wd\z@
\setbox\@ne\hbox{\m@th$#2$}\ifnum\w@dth<\wd\@ne \w@dth=\wd\@ne \fi
\advance\w@dth by 1.2em}

\def\t@^#1_#2{\allowbreak\def\n@one{#1}\def\n@two{#2}\mathrel
{\setw@dth{#1}{#2} \mathop{\hbox to
\w@dth{\rightarrowfill}}\limits \ifx\n@one\empty\else
^{\box\z@}\fi \ifx\n@two\empty\else _{\box\@ne}\fi}}
\def\t@@^#1{\@ifnextchar_{\t@^{#1}}{\t@^{#1}_{}}}
\def\to{\@ifnextchar^{\t@@}{\t@@^{}}}

\def\t@left^#1_#2{\def\n@one{#1}\def\n@two{#2}\mathrel{\setw@dth{#1}{#2}
\mathop{\hbox to \w@dth{\leftarrowfill}}\limits
\ifx\n@one\empty\else ^{\box\z@}\fi \ifx\n@two\empty\else
_{\box\@ne}\fi}}
\def\t@@left^#1{\@ifnextchar_{\t@left^{#1}}{\t@left^{#1}_{}}}
\def\toleft{\@ifnextchar^{\t@@left}{\t@@left^{}}}

\def\two@^#1_#2{\allowbreak
\def\n@one{#1}\def\n@two{#2}\mathrel{\setw@dth{#1}{#2}
\mathop{\vcenter{\lineskip\z@\baselineskip\z@
                 \hbox to \w@dth{\rightarrowfill}%
                 \hbox to \w@dth{\rightarrowfill}}%
       }\limits
\ifx\n@one\empty\else ^{\box\z@}\fi \ifx\n@two\empty\else
_{\box\@ne}\fi}}
\def\tw@@^#1{\@ifnextchar _{\two@^{#1}}{\two@^{#1}_{}}}
\def\two{\@ifnextchar ^{\tw@@}{\tw@@^{}}}

\def\tofr@^#1_#2{\def\n@one{#1}\def\n@two{#2}\mathrel{\setw@dth{#1}{#2}
\mathop{\vcenter{\hbox to \w@dth{\rightarrowfill}\kern-1.7ex
                 \hbox to \w@dth{\leftarrowfill}}%
       }\limits
\ifx\n@one\empty\else ^{\box\z@}\fi \ifx\n@two\empty\else
_{\box\@ne}\fi}}
\def\t@fr@^#1{\@ifnextchar_ {\tofr@^{#1}}{\tofr@^{#1}_{}}}
\def\tofro{\@ifnextchar^ {\t@fr@}{\t@fr@^{}}}

\def\mon{\mathop{\m@th\hbox to
      14.6\P@{\lasyb\char'51\hskip-2.1\P@$\arrext$\hss
$\mathord\rightarrow$}}\limits} 
\def\leftmono{\mathrel{\m@th\hbox to
14.6\P@{$\mathord\leftarrow$\hss$\arrext$\hskip-2.1\P@\lasyb\char'50%
}}\limits} 
\mathchardef\arrext="0200       

\def\settypes(#1,#2,#3){\arrowtypea#1 \arrowtypeb#2 \arrowtypec#3}
\def\settoheight#1#2{\setbox\@tempboxa\hbox{#2}#1\ht\@tempboxa\relax}%
\def\settodepth#1#2{\setbox\@tempboxa\hbox{#2}#1\dp\@tempboxa\relax}%
\def\settokens`#1`#2`#3`#4`{%
     \def\tokena{#1}\def\tokenb{#2}\def\tokenc{#3}\def\tokend{#4}}
\def\setsqparms[#1`#2`#3`#4;#5`#6]{%
\arrowtypea #1 \arrowtypeb #2 \arrowtypec #3 \arrowtyped #4
\width #5 \height #6 }
\def\setpos(#1,#2){\xpos=#1 \ypos#2}

\def\settriparms[#1`#2`#3;#4]{\settripairparms[#1`#2`#3`1`1;#4]}%

\def\settripairparms[#1`#2`#3`#4`#5;#6]{%
\arrowtypea #1 \arrowtypeb #2 \arrowtypec #3 \arrowtyped #4
\arrowtypee #5 \width #6 \height #6 }

\def\resetparms{\settripairparms[1`1`1`1`1;500]\width 500}

\resetparms

\def\mvector(#1,#2)#3{
\put(0,0){\vector(#1,#2){#3}}%
\put(0,0){\vector(#1,#2){26}}%
}
\def\evector(#1,#2)#3{{
\arrowlength #3
\put(0,0){\vector(#1,#2){\arrowlength}}%
\advance \arrowlength by-30
\put(0,0){\vector(#1,#2){\arrowlength}}%
}}

\def\horsize#1#2{%
\settowidth{\tempdimen}{$#2$}%
#1=\tempdimen \divide #1 by\unitlength }

\def\vertsize#1#2{%
\settoheight{\tempdimen}{$#2$}%
#1=\tempdimen
\settodepth{\tempdimen}{$#2$}%
\advance #1 by\tempdimen \divide #1 by\unitlength }

\def\putvector(#1,#2)(#3,#4)#5#6{{%
\ifnum3<\arrowtype \putdashvector(#1,#2)(#3,#4)#5\arrowtype \else
\ifnum\arrowtype<-3 \putdashvector(#1,#2)(#3,#4)#5\arrowtype \else
\xpos=#1 \ypos=#2 \run=#3 \rise=#4 \arrowlength=#5 \ifnum
\arrowtype<0
    \ifnum \run=0
        \advance \ypos by-\arrowlength
    \else
        \tempcounta \arrowlength
        \multiply \tempcounta by\rise
        \divide \tempcounta by\run
        \ifnum\run>0
            \advance \xpos by\arrowlength
            \advance \ypos by\tempcounta
        \else
            \advance \xpos by-\arrowlength
            \advance \ypos by-\tempcounta
        \fi
    \fi
    \multiply \arrowtype by-1
    \multiply \rise by-1
    \multiply \run by-1
\fi \ifcase \arrowtype
\or \put(\xpos,\ypos){\vector(\run,\rise){\arrowlength}}%
\or \put(\xpos,\ypos){\mvector(\run,\rise)\arrowlength}%
\or \put(\xpos,\ypos){\evector(\run,\rise){\arrowlength}}%
\fi\fi\fi }}

\def\putsplitvector(#1,#2)#3#4{
\xpos #1 \ypos #2 \arrowtype #4 \halflength #3 \arrowlength #3
\gap 140 \advance \halflength by-\gap \divide \halflength by2
\ifnum\arrowtype>0
   \ifcase \arrowtype
   \or \put(\xpos,\ypos){\line(0,-1){\halflength}}%
       \advance\ypos by-\halflength
       \advance\ypos by-\gap
       \put(\xpos,\ypos){\vector(0,-1){\halflength}}%
   \or \put(\xpos,\ypos){\line(0,-1)\halflength}%
       \put(\xpos,\ypos){\vector(0,-1)3}%
       \advance\ypos by-\halflength
       \advance\ypos by-\gap
       \put(\xpos,\ypos){\vector(0,-1){\halflength}}%
   \or \put(\xpos,\ypos){\line(0,-1)\halflength}%
       \advance\ypos by-\halflength
       \advance\ypos by-\gap
       \put(\xpos,\ypos){\evector(0,-1){\halflength}}%
   \fi
\else \arrowtype=-\arrowtype
   \ifcase\arrowtype
   \or \advance \ypos by-\arrowlength
       \put(\xpos,\ypos){\line(0,1){\halflength}}%
       \advance\ypos by\halflength
       \advance\ypos by\gap
       \put(\xpos,\ypos){\vector(0,1){\halflength}}%
   \or \advance \ypos by-\arrowlength
       \put(\xpos,\ypos){\line(0,1)\halflength}%
       \put(\xpos,\ypos){\vector(0,1)3}%
       \advance\ypos by\halflength
       \advance\ypos by\gap
       \put(\xpos,\ypos){\vector(0,1){\halflength}}%
   \or \advance \ypos by-\arrowlength
       \put(\xpos,\ypos){\line(0,1)\halflength}%
       \advance\ypos by\halflength
       \advance\ypos by\gap
       \put(\xpos,\ypos){\evector(0,1){\halflength}}%
   \fi
\fi }

\def\putmorphism(#1)(#2,#3)[#4`#5`#6]#7#8#9{{%
\run #2 \rise #3 \ifnum\rise=0
  \puthmorphism(#1)[#4`#5`#6]{#7}{#8}#9%
\else\ifnum\run=0
  \putvmorphism(#1)[#4`#5`#6]{#7}{#8}#9%
\else
\setpos(#1)%
\arrowlength #7 \arrowtype #8 \ifnum\run=0 \else\ifnum\rise=0
\else \ifnum\run>0
    \coefa=1
\else
   \coefa=-1
\fi \ifnum\arrowtype>0
   \coefb=0
   \coefc=-1
\else
   \coefb=\coefa
   \coefc=1
   \arrowtype=-\arrowtype
\fi \width=2 \multiply \width by\run \divide \width by\rise
\ifnum \width<0  \width=-\width\fi \advance\width by60 \if l#9
\width=-\width\fi
\putbox(\xpos,\ypos){#4}
{\multiply \coefa by\arrowlength
\advance\xpos by\coefa \multiply \coefa by\rise \divide \coefa
by\run \advance \ypos by\coefa
\putbox(\xpos,\ypos){#5} }%
{\multiply \coefa by\arrowlength
\divide \coefa by2 \advance \xpos by\coefa \advance \xpos by\width
\multiply \coefa by\rise \divide \coefa by\run \advance \ypos
by\coefa
\if l#9%
   \putrbox(\xpos,\ypos){#6}%
\else\if r#9%
   \putlbox(\xpos,\ypos){#6}%
\fi\fi }%
{\multiply \rise by-\coefc
\multiply \run by-\coefc \multiply \coefb by\arrowlength \advance
\xpos by\coefb \multiply \coefb by\rise \divide \coefb by\run
\advance \ypos by\coefb \multiply \coefc by70 \advance \ypos
by\coefc \multiply \coefc by\run \divide \coefc by\rise \advance
\xpos by\coefc \multiply \coefa by140 \multiply \coefa by\run
\divide \coefa by\rise \advance \arrowlength by\coefa
\ifcase\arrowtype
\or \put(\xpos,\ypos){\vector(\run,\rise){\arrowlength}}%
\or \put(\xpos,\ypos){\mvector(\run,\rise){\arrowlength}}%
\or \put(\xpos,\ypos){\evector(\run,\rise){\arrowlength}}%
\fi}\fi\fi\fi\fi}}

\newcount\numbdashes \newcount\lengthdash \newcount\increment

\def\howmanydashes{
\numbdashes=\arrowlength \lengthdash=40 \divide\numbdashes by
\lengthdash \lengthdash=\arrowlength \divide\lengthdash by
\numbdashes
\increment=\lengthdash \multiply\lengthdash by 3
\divide\lengthdash by 5 }

\def\putdashvector(#1)(#2,#3)#4#5{%
\ifnum#3=0 \putdashhvector(#1){#4}#5 \else \ifnum#2=0
\putdashvvector(#1){#4}#5\fi\fi}

\def\putdashhvector(#1,#2)#3#4{{%
\arrowlength=#3 \howmanydashes
\multiput(#1,#2)(\increment,0){\numbdashes}%
{\vrule height .4pt width \lengthdash\unitlength} \arrowtype=#4
\xpos=#1 \ifnum\arrowtype<0 \advance\arrowtype by 7 \fi
\ifcase\arrowtype \or \advance\xpos by 10
    \put(\xpos,#2){\vector(-1,0){\lengthdash}}
    \advance\xpos by 40
    \put(\xpos,#2){\vector(-1,0){\lengthdash}}
\or \advance \xpos by 10
    \put(\xpos,#2){\vector(-1,0){\lengthdash}}
    \advance\xpos by  \arrowlength
    \advance\xpos by  -50
    \put(\xpos,#2){\vector(-1,0){\lengthdash}}
\or \advance\xpos by 10
    \put(\xpos,#2){\vector(-1,0){\lengthdash}}
\or \advance\xpos by \arrowlength
    \advance\xpos by -\lengthdash
    \put(\xpos,#2){\vector(1,0){\lengthdash}}
\or {\advance\xpos by 10
    \put(\xpos,#2){\vector(1,0){\lengthdash}}}
    \advance\xpos by \arrowlength
    \advance\xpos by -\lengthdash
    \put(\xpos,#2){\vector(1,0){\lengthdash}}
\or \advance\xpos by \arrowlength
    \advance\xpos by -\lengthdash
    \put(\xpos,#2){\vector(1,0){\lengthdash}}
    \advance\xpos by -40
    \put(\xpos,#2){\vector(1,0){\lengthdash}}
   \fi
}}

\def\putdashvvector(#1,#2)#3#4{{%
\arrowlength=#3 \howmanydashes \ypos=#2 \advance\ypos by
-\arrowlength
\multiput(#1,#2)(0,\increment){\numbdashes}%
    {\vrule width .4pt height \lengthdash\unitlength}
\arrowtype=#4 \ypos=#2 \ifnum\arrowtype<0 \advance\arrowtype by 7
\fi \ifcase\arrowtype \or \advance\ypos by \arrowlength
\advance\ypos by -40
    \put(#1,\ypos){\vector(0,1){\lengthdash}}
    \advance\ypos by -40
    \put(#1,\ypos){\vector(0,1){\lengthdash}}
\or \advance\ypos by 10
    \put(#1,\ypos){\vector(0,1){\lengthdash}}
    \advance\ypos by \arrowlength \advance\ypos by -40
    \put(#1,\ypos){\vector(0,1){\lengthdash}}
\or \advance\ypos by \arrowlength \advance\ypos by -40
    \put(#1,\ypos){\vector(0,1){\lengthdash}}
\or \advance\ypos by 10
    \put(#1,\ypos){\vector(0,-1){\lengthdash}}
\or \advance\ypos by 10
    \put(#1,\ypos){\vector(0,-1){\lengthdash}}
    \advance\ypos by \arrowlength \advance\ypos by -40
    \put(#1,\ypos){\vector(0,-1){\lengthdash}}
\or \advance\ypos by 10
    \put(#1,\ypos){\vector(0,-1){\lengthdash}}
    \advance\ypos by 40
    \put(#1,\ypos){\vector(0,-1){\lengthdash}}
\fi }}

\def\puthmorphism(#1,#2)[#3`#4`#5]#6#7#8{{%
\xpos #1 \ypos #2 \width #6 \arrowlength #6 \arrowtype=#7
\putbox(\xpos,\ypos){#3\vphantom{#4}}%
{\advance \xpos by\arrowlength
\putbox(\xpos,\ypos){\vphantom{#3}#4}}%
\horsize{\tempcounta}{#3}%
\horsize{\tempcountb}{#4}%
\divide \tempcounta by2 \divide \tempcountb by2 \advance
\tempcounta by30 \advance \tempcountb by30 \advance \xpos
by\tempcounta \advance \arrowlength by-\tempcounta \advance
\arrowlength by-\tempcountb
\putvector(\xpos,\ypos)(1,0)\arrowlength\arrowtype \divide
\arrowlength by2 \advance \xpos by\arrowlength
\vertsize{\tempcounta}{#5}%
\divide\tempcounta by2 \advance \tempcounta by20
\if a#8 %
   \advance \ypos by\tempcounta
   \putbox(\xpos,\ypos){#5}%
\else
   \advance \ypos by-\tempcounta
   \putbox(\xpos,\ypos){#5}%
\fi}}

\def\putvmorphism(#1,#2)[#3`#4`#5]#6#7#8{{%
\xpos #1 \ypos #2 \arrowlength #6 \arrowtype #7
\settowidth{\xlen}{$#5$}%
\putbox(\xpos,\ypos){#3}%
{\advance \ypos by-\arrowlength
\putbox(\xpos,\ypos){#4}}%
{\advance\arrowlength by-140 \advance \ypos by-70 \ifdim\xlen>0pt
   \if m#8%
      \putsplitvector(\xpos,\ypos)\arrowlength\arrowtype
   \else
   \putvector(\xpos,\ypos)(0,-1)\arrowlength\arrowtype
   \fi
\else
   \putvector(\xpos,\ypos)(0,-1)\arrowlength\arrowtype
\fi}%
\ifdim\xlen>0pt
   \divide \arrowlength by2
   \advance\ypos by-\arrowlength
   \if l#8%
      \advance \xpos by-40
      \putrbox(\xpos,\ypos){#5}%
   \else\if r#8%
      \advance \xpos by40
      \putlbox(\xpos,\ypos){#5}%
   \else
      \putbox(\xpos,\ypos){#5}%
   \fi\fi
\fi }}

\def\putsquarep<#1>(#2)[#3;#4`#5`#6`#7]{{%
\setsqparms[#1]%
\setpos(#2)%
\settokens`#3`%
\puthmorphism(\xpos,\ypos)[\tokenc`\tokend`{#7}]{\width}{\arrowtyped}b%
\advance\ypos by \height
\puthmorphism(\xpos,\ypos)[\tokena`\tokenb`{#4}]{\width}{\arrowtypea}a%
\putvmorphism(\xpos,\ypos)[``{#5}]{\height}{\arrowtypeb}l%
\advance\xpos by \width
\putvmorphism(\xpos,\ypos)[``{#6}]{\height}{\arrowtypec}r%
}}

\def\putsquare{\@ifnextchar <{\putsquarep}{\putsquarep%
   <\arrowtypea`\arrowtypeb`\arrowtypec`\arrowtyped;\width`\height>}}
\def\square{\@ifnextchar< {\squarep}{\squarep
   <\arrowtypea`\arrowtypeb`\arrowtypec`\arrowtyped;\width`\height>}}
\def\squarep<#1>[#2`#3`#4`#5;#6`#7`#8`#9]{{
\setsqparms[#1]
\diagram
\putsquarep<\arrowtypea`\arrowtypeb`\arrowtypec`
\arrowtyped;\width`\height>
(0,0)[#2`#3`#4`{#5};#6`#7`#8`{#9}]
\enddiagram
}}                                                 
\def\putptrianglep<#1>(#2,#3)[#4`#5`#6;#7`#8`#9]{{%
\settriparms[#1]%
\xpos=#2 \ypos=#3 \advance\ypos by \height
\puthmorphism(\xpos,\ypos)[#4`#5`{#7}]{\height}{\arrowtypea}a%
\putvmorphism(\xpos,\ypos)[`#6`{#8}]{\height}{\arrowtypeb}l%
\advance\xpos by\height
\putmorphism(\xpos,\ypos)(-1,-1)[``{#9}]{\height}{\arrowtypec}r%
}}

\def\putptriangle{\@ifnextchar <{\putptrianglep}{\putptrianglep
   <\arrowtypea`\arrowtypeb`\arrowtypec;\height>}}
\def\ptriangle{\@ifnextchar <{\ptrianglep}{\ptrianglep
   <\arrowtypea`\arrowtypeb`\arrowtypec;\height>}}
\def\ptrianglep<#1>[#2`#3`#4;#5`#6`#7]{{
\settriparms[#1]
\diagram
\putptrianglep<\arrowtypea`\arrowtypeb`
\arrowtypec;\height>
(0,0)[#2`#3`#4;#5`#6`{#7}]
\enddiagram
}}                                            

\def\putqtrianglep<#1>(#2,#3)[#4`#5`#6;#7`#8`#9]{{%
\settriparms[#1]%
\xpos=#2 \ypos=#3 \advance\ypos by\height
\puthmorphism(\xpos,\ypos)[#4`#5`{#7}]{\height}{\arrowtypea}a%
\putmorphism(\xpos,\ypos)(1,-1)[``{#8}]{\height}{\arrowtypeb}l%
\advance\xpos by\height
\putvmorphism(\xpos,\ypos)[`#6`{#9}]{\height}{\arrowtypec}r%
}}

\def\putqtriangle{\@ifnextchar <{\putqtrianglep}{\putqtrianglep
   <\arrowtypea`\arrowtypeb`\arrowtypec;\height>}}
\def\qtriangle{\@ifnextchar <{\qtrianglep}{\qtrianglep
   <\arrowtypea`\arrowtypeb`\arrowtypec;\height>}}
\def\qtrianglep<#1>[#2`#3`#4;#5`#6`#7]{{
\settriparms[#1]
\width=\height                                
\diagram
\putqtrianglep<\arrowtypea`\arrowtypeb`
\arrowtypec;\height>
(0,0)[#2`#3`#4;#5`#6`{#7}]
\enddiagram
}}

\def\putdtrianglep<#1>(#2,#3)[#4`#5`#6;#7`#8`#9]{{%
\settriparms[#1]%
\xpos=#2 \ypos=#3
\puthmorphism(\xpos,\ypos)[#5`#6`{#9}]{\height}{\arrowtypec}b%
\advance\xpos by \height \advance\ypos by\height
\putmorphism(\xpos,\ypos)(-1,-1)[``{#7}]{\height}{\arrowtypea}l%
\putvmorphism(\xpos,\ypos)[#4``{#8}]{\height}{\arrowtypeb}r%
}}

\def\putdtriangle{\@ifnextchar <{\putdtrianglep}{\putdtrianglep
   <\arrowtypea`\arrowtypeb`\arrowtypec;\height>}}
\def\dtriangle{\@ifnextchar <{\dtrianglep}{\dtrianglep
   <\arrowtypea`\arrowtypeb`\arrowtypec;\height>}}
\def\dtrianglep<#1>[#2`#3`#4;#5`#6`#7]{{
\settriparms[#1]
\width=\height                                
\diagram
\putdtrianglep<\arrowtypea`\arrowtypeb`
\arrowtypec;\height>
(0,0)[#2`#3`#4;#5`#6`{#7}]
\enddiagram
}}

\def\putbtrianglep<#1>(#2,#3)[#4`#5`#6;#7`#8`#9]{{%
\settriparms[#1]%
\xpos=#2 \ypos=#3
\puthmorphism(\xpos,\ypos)[#5`#6`{#9}]{\height}{\arrowtypec}b%
\advance\ypos by\height
\putmorphism(\xpos,\ypos)(1,-1)[``{#8}]{\height}{\arrowtypeb}r%
\putvmorphism(\xpos,\ypos)[#4``{#7}]{\height}{\arrowtypea}l%
}}

\def\putbtriangle{\@ifnextchar <{\putbtrianglep}{\putbtrianglep
   <\arrowtypea`\arrowtypeb`\arrowtypec;\height>}}
\def\btriangle{\@ifnextchar <{\btrianglep}{\btrianglep
   <\arrowtypea`\arrowtypeb`\arrowtypec;\height>}}
\def\btrianglep<#1>[#2`#3`#4;#5`#6`#7]{{
\settriparms[#1]
\width=\height                               
\diagram
\putbtrianglep<\arrowtypea`\arrowtypeb`
\arrowtypec;\height>
(0,0)[#2`#3`#4;#5`#6`{#7}]
\enddiagram
}}

\def\putAtrianglep<#1>(#2,#3)[#4`#5`#6;#7`#8`#9]{{%
\settriparms[#1]%
\xpos=#2 \ypos=#3 {\multiply \height by2
\puthmorphism(\xpos,\ypos)[#5`#6`{#9}]{\height}{\arrowtypec}b}%
\advance\xpos by\height \advance\ypos by\height
\putmorphism(\xpos,\ypos)(-1,-1)[#4``{#7}]{\height}{\arrowtypea}l%
\putmorphism(\xpos,\ypos)(1,-1)[``{#8}]{\height}{\arrowtypeb}r%
}}

\def\putAtriangle{\@ifnextchar <{\putAtrianglep}{\putAtrianglep
   <\arrowtypea`\arrowtypeb`\arrowtypec;\height>}}
\def\Atriangle{\@ifnextchar <{\Atrianglep}{\Atrianglep
   <\arrowtypea`\arrowtypeb`\arrowtypec;\height>}}
\def\Atrianglep<#1>[#2`#3`#4;#5`#6`#7]{{
\settriparms[#1]
\width=\height                                     
\diagram
\putAtrianglep<\arrowtypea`\arrowtypeb`
\arrowtypec;\height>
(0,0)[#2`#3`#4;#5`#6`{#7}]
\enddiagram
}}

\def\putAtrianglepairp<#1>(#2)[#3;#4`#5`#6`#7`#8]{{%
\settripairparms[#1]%
\setpos(#2)%
\settokens`#3`%
\puthmorphism(\xpos,\ypos)[\tokenb`\tokenc`{#7}]{\height}{\arrowtyped}b%
\advance\xpos by\height
\puthmorphism(\xpos,\ypos)[\phantom{\tokenc}`\tokend`{#8}]%
{\height}{\arrowtypee}b%
\advance\ypos by\height
\putmorphism(\xpos,\ypos)(-1,-1)[\tokena``{#4}]{\height}{\arrowtypea}l%
\putvmorphism(\xpos,\ypos)[``{#5}]{\height}{\arrowtypeb}m%
\putmorphism(\xpos,\ypos)(1,-1)[``{#6}]{\height}{\arrowtypec}r%
}}

\def\putAtrianglepair{\@ifnextchar <{\putAtrianglepairp}{\putAtrianglepairp%
   <\arrowtypea`\arrowtypeb`\arrowtypec`\arrowtyped`\arrowtypee;\height>}}
\def\Atrianglepair{\@ifnextchar <{\Atrianglepairp}{\Atrianglepairp%
   <\arrowtypea`\arrowtypeb`\arrowtypec`\arrowtyped`\arrowtypee;\height>}}

\def\Atrianglepairp<#1>[#2;#3`#4`#5`#6`#7]{{
\settripairparms[#1]
\settokens`#2`
\width=\height                                
\diagram
\putAtrianglepairp                            
<\arrowtypea`\arrowtypeb`\arrowtypec`
\arrowtyped`\arrowtypee;\height>
(0,0)[{#2};#3`#4`#5`#6`{#7}]
\enddiagram
}}

\def\putVtrianglep<#1>(#2,#3)[#4`#5`#6;#7`#8`#9]{{%
\settriparms[#1]%
\xpos=#2 \ypos=#3 \advance\ypos by\height {\multiply\height by2
\puthmorphism(\xpos,\ypos)[#4`#5`{#7}]{\height}{\arrowtypea}a}%
\putmorphism(\xpos,\ypos)(1,-1)[`#6`{#8}]{\height}{\arrowtypeb}l%
\advance\xpos by\height \advance\xpos by\height
\putmorphism(\xpos,\ypos)(-1,-1)[``{#9}]{\height}{\arrowtypec}r%
}}

\def\putVtriangle{\@ifnextchar <{\putVtrianglep}{\putVtrianglep
   <\arrowtypea`\arrowtypeb`\arrowtypec;\height>}}
\def\Vtriangle{\@ifnextchar <{\Vtrianglep}{\Vtrianglep
   <\arrowtypea`\arrowtypeb`\arrowtypec;\height>}}
\def\Vtrianglep<#1>[#2`#3`#4;#5`#6`#7]{{
\settriparms[#1]
\width=\height                                 
\diagram
\putVtrianglep<\arrowtypea`\arrowtypeb`
\arrowtypec;\height>
(0,0)[#2`#3`#4;#5`#6`{#7}]
\enddiagram
}}

\def\putVtrianglepairp<#1>(#2)[#3;#4`#5`#6`#7`#8]{{
\settripairparms[#1]%
\setpos(#2)%
\settokens`#3`%
\advance\ypos by\height
\putmorphism(\xpos,\ypos)(1,-1)[`\tokend`{#6}]{\height}{\arrowtypec}l%
\puthmorphism(\xpos,\ypos)[\tokena`\tokenb`{#4}]{\height}{\arrowtypea}a%
\advance\xpos by\height
\puthmorphism(\xpos,\ypos)[\phantom{\tokenb}`\tokenc`{#5}]%
{\height}{\arrowtypeb}a%
\putvmorphism(\xpos,\ypos)[``{#7}]{\height}{\arrowtyped}m%
\advance\xpos by\height
\putmorphism(\xpos,\ypos)(-1,-1)[``{#8}]{\height}{\arrowtypee}r%
}}

\def\putVtrianglepair{\@ifnextchar <{\putVtrianglepairp}{\putVtrianglepairp%
    <\arrowtypea`\arrowtypeb`\arrowtypec`\arrowtyped`\arrowtypee;\height>}}
\def\Vtrianglepair{\@ifnextchar <{\Vtrianglepairp}{\Vtrianglepairp%
    <\arrowtypea`\arrowtypeb`\arrowtypec`\arrowtyped`\arrowtypee;\height>}}
\def\Vtrianglepairp<#1>[#2;#3`#4`#5`#6`#7]{{
\settripairparms[#1]
\settokens`#2`
\diagram
\putVtrianglepairp                             
<\arrowtypea`\arrowtypeb`\arrowtypec`
\arrowtyped`\arrowtypee;\height>
(0,0)[{#2};#3`#4`#5`#6`{#7}]
\enddiagram
}}

\def\putCtrianglep<#1>(#2,#3)[#4`#5`#6;#7`#8`#9]{{%
\settriparms[#1]%
\xpos=#2 \ypos=#3 \advance\ypos by\height
\putmorphism(\xpos,\ypos)(1,-1)[``{#9}]{\height}{\arrowtypec}l%
\advance\xpos by\height \advance\ypos by\height
\putmorphism(\xpos,\ypos)(-1,-1)[#4`#5`{#7}]{\height}{\arrowtypea}l%
{\multiply\height by 2
\putvmorphism(\xpos,\ypos)[`#6`{#8}]{\height}{\arrowtypeb}r}%
}}

\def\putCtriangle{\@ifnextchar <{\putCtrianglep}{\putCtrianglep
    <\arrowtypea`\arrowtypeb`\arrowtypec;\height>}}
\def\Ctriangle{\@ifnextchar <{\Ctrianglep}{\Ctrianglep
    <\arrowtypea`\arrowtypeb`\arrowtypec;\height>}}
\def\Ctrianglep<#1>[#2`#3`#4;#5`#6`#7]{{
\settriparms[#1]
\width=\height                               
\diagram
\putCtrianglep<\arrowtypea`\arrowtypeb`
\arrowtypec;\height>
(0,0)[#2`#3`#4;#5`#6`{#7}]
\enddiagram
}}                                           
\def\putDtrianglep<#1>(#2,#3)[#4`#5`#6;#7`#8`#9]{{%
\settriparms[#1]%
\xpos=#2 \ypos=#3 \advance\xpos by\height \advance\ypos by\height
\putmorphism(\xpos,\ypos)(-1,-1)[``{#9}]{\height}{\arrowtypec}r%
\advance\xpos by-\height \advance\ypos by\height
\putmorphism(\xpos,\ypos)(1,-1)[`#5`{#8}]{\height}{\arrowtypeb}r%
{\multiply\height by 2
\putvmorphism(\xpos,\ypos)[#4`#6`{#7}]{\height}{\arrowtypea}l}%
}}

\def\putDtriangle{\@ifnextchar <{\putDtrianglep}{\putDtrianglep
    <\arrowtypea`\arrowtypeb`\arrowtypec;\height>}}
\def\Dtriangle{\@ifnextchar <{\Dtrianglep}{\Dtrianglep
   <\arrowtypea`\arrowtypeb`\arrowtypec;\height>}}
\def\Dtrianglep<#1>[#2`#3`#4;#5`#6`#7]{{
\settriparms[#1]
\width=\height                              
\diagram
\putDtrianglep<\arrowtypea`\arrowtypeb`
\arrowtypec;\height>
(0,0)[#2`#3`#4;#5`#6`{#7}]
\enddiagram
}}                                          
\def\setrecparms[#1`#2]{\width=#1 \height=#2}%

\def\recursep<#1`#2>[#3;#4`#5`#6`#7`#8]{{\m@th
\width=#1 \height=#2 \settokens`#3`
\settowidth{\tempdimen}{$\tokena$} \ifdim\tempdimen=0pt
  \savebox{\tempboxa}{\hbox{$\tokenb$}}%
  \savebox{\tempboxb}{\hbox{$\tokend$}}%
  \savebox{\tempboxc}{\hbox{$#6$}}%
\else
  \savebox{\tempboxa}{\hbox{$\hbox{$\tokena$}\times\hbox{$\tokenb$}$}}%
  \savebox{\tempboxb}{\hbox{$\hbox{$\tokena$}\times\hbox{$\tokend$}$}}%
  \savebox{\tempboxc}{\hbox{$\hbox{$\tokena$}\times\hbox{$#6$}$}}%
\fi \ypos=\height \divide\ypos by 2 \xpos=\ypos \advance\xpos by
\width \bfig
\putCtrianglep<-1`1`1;\ypos>(0,0)[`\tokenc`;#5`#6`{#7}]%
\puthmorphism(\ypos,0)[\tokend`\usebox{\tempboxb}`{#8}]{\width}{-1}b%
\puthmorphism(\ypos,\height)[\tokenb`\usebox{\tempboxa}`{#4}]{\width}{-1}a%
\advance\ypos by \width
\putvmorphism(\ypos,\height)[``\usebox{\tempboxc}]{\height}1r%
\efig }}

\def\recurse{\@ifnextchar <{\recursep}{\recursep<\width`\height>}}

\def\puttwohmorphisms(#1,#2)[#3`#4;#5`#6]#7#8#9{{%
\puthmorphism(#1,#2)[#3`#4`]{#7}0a \ypos=#2 \advance\ypos by 20
\puthmorphism(#1,\ypos)[\phantom{#3}`\phantom{#4}`#5]{#7}{#8}a
\advance\ypos by -40
\puthmorphism(#1,\ypos)[\phantom{#3}`\phantom{#4}`#6]{#7}{#9}b }}

\def\puttwovmorphisms(#1,#2)[#3`#4;#5`#6]#7#8#9{{%
\putvmorphism(#1,#2)[#3`#4`]{#7}0a \xpos=#1 \advance\xpos by -20
\putvmorphism(\xpos,#2)[\phantom{#3}`\phantom{#4}`#5]{#7}{#8}l
\advance\xpos by 40
\putvmorphism(\xpos,#2)[\phantom{#3}`\phantom{#4}`#6]{#7}{#9}r }}

\def\puthcoequalizer(#1)[#2`#3`#4;#5`#6`#7]#8#9{{%
\setpos(#1)%
\puttwohmorphisms(\xpos,\ypos)[#2`#3;#5`#6]{#8}11%
\advance\xpos by #8
\puthmorphism(\xpos,\ypos)[\phantom{#3}`#4`#7]{#8}1{#9} }}

\def\putvcoequalizer(#1)[#2`#3`#4;#5`#6`#7]#8#9{{%
\setpos(#1)%
\puttwovmorphisms(\xpos,\ypos)[#2`#3;#5`#6]{#8}11%
\advance\ypos by -#8
\putvmorphism(\xpos,\ypos)[\phantom{#3}`#4`#7]{#8}1{#9} }}

\def\putthreehmorphisms(#1)[#2`#3;#4`#5`#6]#7(#8)#9{{%
\setpos(#1) \settypes(#8)
\if a#9 %
     \vertsize{\tempcounta}{#5}%
     \vertsize{\tempcountb}{#6}%
     \ifnum \tempcounta<\tempcountb \tempcounta=\tempcountb \fi
\else
     \vertsize{\tempcounta}{#4}%
     \vertsize{\tempcountb}{#5}%
     \ifnum \tempcounta<\tempcountb \tempcounta=\tempcountb \fi
\fi \advance \tempcounta by 60
\puthmorphism(\xpos,\ypos)[#2`#3`#5]{#7}{\arrowtypeb}{#9}
\advance\ypos by \tempcounta
\puthmorphism(\xpos,\ypos)[\phantom{#2}`\phantom{#3}`#4]{#7}{\arrowtypea}{#9}
\advance\ypos by -\tempcounta \advance\ypos by -\tempcounta
\puthmorphism(\xpos,\ypos)[\phantom{#2}`\phantom{#3}`#6]{#7}{\arrowtypec}{#9}
}}

\def\setarrowtoks[#1`#2`#3`#4`#5`#6]{%
\def\toka{#1}
\def\tokb{#2}
\def\tokc{#3}
\def\tokd{#4}
\def\toke{#5}
\def\tokf{#6}
}
\def\hex{\@ifnextchar <{\hexp}{\hexp<1000`400>}}
\def\hexp<#1`#2>[#3`#4`#5`#6`#7`#8;#9]{%
\setarrowtoks[#9] \yext=#2 \advance \yext by #2 \xext=#1
\advance\xext by \yext \bfig
\putCtriangle<-1`0`1;#2>(0,0)[`#5`;\tokb``\tokd] \xext=#1
\yext=#2 \advance \yext by #2
\putsquare<1`0`0`1;\xext`\yext>(#2,0)[#3`#4`#7`#8;\toka```\tokf]
\advance \xext by #2
\putDtriangle<0`1`-1;#2>(\xext,0)[`#6`;`\tokc`\toke] \efig }

\makeatother